\documentclass[12pt]{article}
\usepackage[paper=a4paper,margin=1in]{geometry}
\usepackage{t1enc}      

\usepackage{amsmath}
\usepackage{amsfonts}
\usepackage{amssymb}
\usepackage{amsthm}
\usepackage{graphicx}
\usepackage{mathrsfs}  

\newcommand{\edth} {\mbox{\symbol{'360}}}

\newcommand{\bi}{\bf i}
\newcommand{\bj}{\bf j}

\newtheorem*{theorem*}{Theorem}

\providecommand{\keywords}[1]
{\small	\textbf{\textit{Keywords:}} #1 }

\numberwithin{equation}{section}

\begin{document}
\bibliographystyle{unsrt}

\title{Three-space from quantum mechanics\footnote{Dedicated to Andr\'as A. 
M\'arton, who is able to transform geometry, physics and philosophy into 
poetry: www.martonaandras.hu }}
\author{L\'aszl\'o B. Szabados \\
Wigner Research Centre for Physics, \\
H-1525 Budapest 114, P. O. Box 49, EU \\
E-mail: lbszab@rmki.kfki.hu}

\maketitle

\begin{abstract}
The spin geometry theorem of Penrose is extended from $SU(2)$ to $E(3)$ 
(Euclidean) invariant elementary quantum mechanical systems. Using the 
natural decomposition of the total angular momentum into its spin and 
orbital parts, the \emph{distance} between the centre-of-mass lines of the 
elementary subsystems of a classical composite system can be recovered 
from their \emph{relative orbital angular momenta} by $E(3)$-invariant 
classical observables. Motivated by this observation, an expression for 
the `empirical distance' between the elementary subsystems of a 
\emph{composite quantum mechanical system}, given in terms of 
$E(3)$-invariant quantum observables, is suggested. It is shown that, in 
the classical limit, this expression reproduces the \emph{a priori} 
Euclidean distance between the subsystems, though at the quantum level 
it has a discrete character. `Empirical' angles and 3-volume elements are 
also considered. 
\end{abstract}



\section{Introduction}
\label{sec-1}

In 1966, Penrose suggested the so-called $SU(2)$ spin network as a simple 
model for a quantum spacetime (which was published much later only in 
\cite{Pe79}, see also \cite{Pe}). In this model, \emph{angles} between 
elementary quantum mechanical subsystems of the Universe were expressed in 
terms of $SU(2)$ \emph{Casimir invariants}, and, using 
combinatorial/graphical techniques, he showed that, in the classical limit, 
these angles tend to angles between directions in the Euclidean 3-space. 
The significance of this result is that the (conformal structure of the) 
`physical 3-space' that we use as an \emph{a priori} given `arena' in which 
the physical objects are thought to be arranged and the interactions between 
them occur is \emph{determined by the quantum physical systems themselves 
in the classical limit}. Later, this result was derived in \cite{Mo} (and 
recently in \cite{Sz21c}) in the usual framework of quantum mechanics; and, 
following \cite{Mo}, it became known as the Spin Geometry Theorem. 

In \cite{Mo}, Moussouris investigated the structure of networks with other 
symmetry groups, in particular with the Poincar\'e group, and derived 
relations between various observables of such systems. However, as far as 
we know, recovering the \emph{distance}, i.e. the \emph{metric} (rather than 
only the \emph{conformal structure} of the Euclidean 3-space/Minkowski vector 
space) from these models is still lacking. 

According to the traditional view, the distance in a metric space is 
associated with a pair of \emph{points}. Thus, the points are usually 
considered to be the primary concept. Nevertheless, from a positivistic, or 
rather a \emph{Machian} view, the position in physical 3-space/spacetime in 
itself is meaningless, and it seems reasonable to speak only about 
\emph{relative positions} of different elementary subsystems of the Universe. 
This view might suggest to consider the distance, or, in general the 
\emph{relations} between points, to be more fundamental than the points; and 
that the points themselves should be \emph{defined} by these relations. 

In classical general relativity the spacetime geometry, and, in particular, 
the distance between the spacetime points, is determined by standard clocks, 
light rays and mirrors. However, at the fundamental level no such instruments 
exist and, also, the light rays are formed by infinitely many photons in 
special configurations. Thus, the \emph{instrumentalist} approach of defining 
the spacetime and its properties, using e.g. some form of `quantum clocks' 
(see e.g. \cite{SaWi}) at the \emph{fundamental level}, does not seem to 
work. 

The philosophy of the present investigations is the positivistic, Machian 
one of \cite{Pe79,Pe} (that we also adopted in \cite{Sz21c}), even from two 
points of view: first, no \emph{a priori} notion of `physical 3-space' or 
`spacetime' is used. These notions should be \emph{defined} in an 
operational way using only existing (material) quantum subsystems of the 
Universe. Second, we speak about angle and distance \emph{only between two 
(elementary) subsystems} of the Universe, without introducing `directions' 
and `points' at all, e.g. in the form of some `position operator'. (For a 
well readable summary of the approaches based on the position operators, 
see \cite{Bac}.) 

A potentially viable mathematical realization of such a program could be 
based on the use of the \emph{algebra of (basic) quantum observables} in 
the \emph{algebraic formulation} of quantum theory. We consider the quantum 
system to be specified completely if its \emph{algebra of (basic) quantum 
observables} (and, if needed, its representation, too) is given. We think 
that the notion of space(time) and all of its properties should be 
\emph{defined} in terms of the \emph{observables} of \emph{elementary} 
systems. Roughly speaking, an \emph{elementary} quantum system is a system 
whose observables are the self-adjoint elements of the \emph{universal 
enveloping algebra} of some `small' Lie algebra, mostly of the symmetry 
group acting on the system, and the states of the system belong to the 
carrier space of some of its unitary, irreducible representations. From 
our point of view, $su(2)$ and the Lie algebras $e(3)$ and $e(1,3)$ of the 
Euclidean and the Poincar\'e group, $E(3)$ and $E(1,3)$, respectively, are 
particularly interesting. (This general idea of the `$G$-invariant 
\emph{elementary} quantum mechanical systems' is motivated by that of the 
Poincar\'e invariant systems of Newton and Wigner \cite{NeWign}.) The 
observables belonging to this `small' Lie algebra will be called the 
\emph{basic observables}. All the structures \emph{on the algebra of 
observables} may be used in the construction, but additional extra 
structures \emph{may not}. In particular, in the present investigations we 
use only the general \emph{kinematical framework} of quantum mechanics, 
but e.g. no evolution equation for the quantum states. 

The above strategy yields apparently paradoxical results: we have well 
defined distance between `points' (or rather straight lines, as we will see) 
that are \emph{not} defined at all in the quantum theory, just like the 
angles between `directions' that are not defined either. The `true' 
geometry of the physical 3-space/spacetime should ultimately be synthesized 
from the `empirical' distances, angles, etc. introduced via the quantum 
observables. 

As a first step in this strategy, in \cite{Sz21c} we gave a simple proof of 
(a version of) the Spin Geometry Theorem by showing how the \emph{conformal 
structure} of the Euclidean 3-space can be recovered in this algebraic 
framework from the basic quantum observables of $SU(2)$-invariant systems. 
The present paper provides one more step: we extend the Spin Geometry 
Theorem to recover the \emph{metric} of the Euclidean 3-space by extending 
the symmetry group from $SU(2)$ to the quantum mechanical Euclidean group 
$E(3)$. Here, by \emph{quantum mechanical} Euclidean group we mean the 
semi-direct product of $SU(2)$ and the group of translations in the 
Euclidean 3-space, rather than the \emph{classical} $E(3)$, which is the 
semi-direct product of $SO(3)$ and the group of translations. 

Technically, the key idea is that the spin part of the total angular 
momentum of a \emph{composite} system is built from the spin of the 
constituent subsystems and their \emph{relative orbital angular momentum}; 
and the latter, in the traditional formulation, contains information on the 
distance between the subsystems. (In fact, already in the last but one 
sentence of \cite{Pe79}, Penrose raised this possibility in 
Poincar\'e-invariant systems.) This idea can in fact be used to define 
the distance between the elementary Poincar\'e-invariant quantum mechanical 
systems as well. However, since the latter is technically different and 
considerably more complicated, results of those investigations will be 
published separately. This distance, expressed by $E(3)$-invariant 
classical observables of the composite system, is our key notion, and is 
called the \emph{empirical distance}. The analogous notion in the quantum 
theory, defined exclusively in terms of the \emph{$E(3)$-invariant basic 
quantum observables}, is motivated by this classical expression. 

The main result of the present paper is that although in quantum theory 
this empirical distance has some discrete and highly non-Euclidean character, 
but in the classical limit it reproduces the \emph{a priori} Euclidean 
distance between the classical centre-of-mass lines of 
point particles, i.e. of the straight lines in $\mathbb{R}^3$. This can be 
considered to be an extension of Penrose's Spin Geometry Theorem from 
$SU(2)$ to $E(3)$-invariant systems. Therefore, the \emph{metric structure} 
of the Euclidean 3-space can be recovered from the quantum theory in the 
classical limit. Note, however, that in the present approach it is 
\emph{the straight lines} rather than the points that emerged as the 
elementary objects in $\mathbb{R}^3$. This feature of the present approach 
is analogous to that of \emph{twistor theory} \cite{PeMacC72,HT,PR2}, where, 
classically, the elementary objects are the (in general twisting) \emph{null 
lines} in Minkowski spacetime with given vorticity or `twist'. The quantity 
here that is analogous to twist is the spin of the elementary systems. Also, 
at the quantum level, $E(3)$-invariant `empirical angles' and `empirical 
3-volume elements' are suggested, and it is shown that they reproduce the 
angles and 3-volume elements in $\mathbb{R}^3$ in the classical limit. 

In the next section we show how the distance of the centre-of-mass lines 
of $E(3)$-invariant elementary \emph{classical} mechanical systems can be 
expressed by their $E(3)$-invariant classical observables. This expression 
will provide the basis of our empirical distance in the quantum theory. In 
Section \ref{sec-3}, we define $E(3)$-invariant elementary quantum mechanical 
systems and summarize their key properties. In particular, we determine 
their centre-of-mass states, which are the closest analogs of those of the 
classical systems, which turn out to be given just by the spin weighted 
spherical harmonics. Section \ref{sec-4} is devoted to the calculation of 
the empirical distance, angles and 3-volume elements. We clarify their 
classical limit there. Some final remarks are given in Section \ref{sec-5}. 
The paper is concluded with appendices, in which certain technical details 
that we used in the main part of the paper are presented. 

In deriving the results we use complex techniques developed in general 
relativity. The related ideas, notations and conventions are mostly those 
of \cite{HT,PR1}, except that the signature of the spatial 3-metric is 
\emph{positive}, rather than negative definite. We do not use abstract 
indices. 


\section{$E(3)$-invariant classical systems}
\label{sec-2}

\subsection{The definition of the elementary systems}
\label{sub-2.1}

A physical system will be called an $E(3)$-invariant \emph{elementary} 
classical mechanical system if its states can be characterized 
\emph{completely} by its linear momentum $p^a$ and angular momentum $J^{ab}$, 
$a,b,...=1,2,3$; under the action of $SO(3)$, they transform as a vector 
and anti-symmetric tensor, respectively, and, under the translation by 
$\xi^a\in\mathbb{R}^3$, as $(p^a,J^{ab})\mapsto(\tilde p^a,\tilde J^{ab})
:=(p^a,J^{ab}+\xi^ap^b-\xi^bp^a)$. Then 
\begin{equation}
P^2:=\delta_{ab}p^ap^b, \hskip 30pt W:=\frac{1}{2}J^{ab}\varepsilon_{abc}p^c
\label{eq:2.1.1}
\end{equation}
are invariant with respect to these transformations. Here $\varepsilon_{abc}$ 
is the Levi-Civita alternating symbol, and $\delta_{ab}$ is the Kronecker 
delta. The space of the linear and angular momenta, endowed with the Lie 
products 
\begin{eqnarray}
&{}&\{p^a,p^b\}=0, \hskip 20pt
  \{p^a,J^{bc}\}=\delta^{ab}p^c-\delta^{ac}p^b, \label{eq:2.1.2a} \\
&{}&\{J^{ab},J^{cd}\}=\delta^{bc}J^{ad}-\delta^{bd}J^{ac}+\delta^{ad}J^{bc}
  -\delta^{ac}J^{bd} \label{eq:2.1.2c}
\end{eqnarray}
form the Lie algebra $e(3)$ of $E(3)$. Then $P^2$ and $W$ have vanishing 
Lie bracket both with $p^a$ and $J^{ab}$, i.e. they are Casimir invariants. 
Lowering and raising the Latin indices will be defined by $\delta_{ab}$ and 
its inverse, respectively.\footnote{Strictly speaking, the abstract Lie 
algebra $e(3)$ is the semi-direct sum of a 3-dimensional commutative ideal 
and $so(3)\approx su(2)$; and the linear and angular momenta belong to 
the \emph{dual space} of these sub-Lie algebras. Hence, the \emph{natural 
positive definite metric} $\delta_{ab}$ on the commutative ideal (coming 
from the Killing--Cartan metric on $so(3)$ via the $SU(2)$ group action on 
it) yields the metric $\delta_{ab}$ on the classical momentum space and the 
whole tensor algebra over this space. Also, the \emph{natural volume 3-form} 
$\varepsilon_{abc}$ on the momentum 3-space comes from the natural volume 
3-form on the ideal. The transformation properties of the linear and angular 
momenta follow from the $E(3)$ multiplication laws. Although, formally, the 
dual $e(3)^*$ of $e(3)$ is not Lie algebra, but it has a natural (linear) 
Poisson manifold structure with the bracket operation 
(\ref{eq:2.1.2a})-(\ref{eq:2.1.2c}). With this structure $e(3)^*$ is 
isomorphic with the Lie algebra $e(3)$. Thus, \emph{all} these 
algebraic/geometric structures are manifestations of those on the 
\emph{abstract} Lie algebra $e(3)$.} 

If $p^a=0$, then $P=0$ and $W=0$. In this degenerate case the subsequent 
strategy to recover the metric structure of $\mathbb{R}^3$ in terms of the 
basic observables does not seem to work. Thus, in the present paper, we 
consider only the $p^a\not=0$ case. 

Next we form $M^a:=J^{ab}p_b$, for which it follows that $M_ap^a=0$, and the 
identity 
\begin{equation}
P^2J^{ab}=\varepsilon^{abc}p_cW+M^ap^b-M^bp^a \label{eq:2.1.3}
\end{equation}
holds. Since the first term on the right of (\ref{eq:2.1.3}) is translation 
invariant while $M^a\mapsto\tilde M^a=M^a+(P^2\delta^a_b-p^ap_b)\xi^b$, this 
identity is usually interpreted as the decomposition of the angular momentum 
to its spin (or rather helicity) and orbital parts, and $M^a$ as ($P^2$-times) 
the centre-of-mass vector of the system. Note that $\delta^a_b-p^ap_b/P^2$ is 
the projection to the 2-plane orthogonal to $p^a$. Thus one can always find 
a 1-parameter family of translations, viz. $\xi^a=-M^a/P^2+up^a$, $u\in
\mathbb{R}$, which yields vanishing centre-of-mass vector, $\tilde M^a=0$. 
The resulting total angular momentum is $\tilde J^{ab}=\varepsilon^{abc}p_cW/
P^2$, which is just the piece of the angular momentum that the Casimir 
invariant $W$ represents. Thus the straight line $q^a(u):=M^a/P^2+up^a$ is 
interpreted as the trajectory of the centre-of-mass point of the system, 
and will be called the system's \emph{centre-of-mass line}. For the sake of 
brevity, we call an elementary system with given $P$ and $W$ a \emph{single 
particle}. 

Since $M_ap^a=0$, the set of the pairs $(p^a,M_a)$ is the cotangent bundle 
$T^*{\cal S}_P$ of the 2-sphere ${\cal S}_P:=\{p^a\in\mathbb{R}^3\vert\, P^2:=
p^ap^b\delta_{ab}={\rm const}\}$ of radius $P$ in the momentum space. Clearly, 
those and only those pairs $(p^a,M_a)$ can be transformed to one another by 
an Euclidean transformation whose Casimir invariants $P$ and $W$ are the same. 
In particular, $(p^a,M_a)$ can always be transformed into $(p^a,0)$ by an 
appropriate translation. This single particle state space, $T^*{\cal S}_P$, 
is homeomorphic to the manifold of the directed straight lines in $\mathbb{R}
^3$: a line $L$ is fixed if its direction and any of its points are specified. 
Now, the direction of $L$ is fixed by the unit vector $p^a/P$, while for a 
point of $L$ we choose the point where $L$ intersects the 2-plane containing 
the origin of $\mathbb{R}^3$ and orthogonal to $p^a/P$. This latter is given 
by $M^a/P^2$. The scale on these lines is fixed by $P$. $W$ is an additional 
structure on $L$: it fixes the component of the total angular momentum vector 
$\frac{1}{2}\varepsilon^{abc}J_{bc}$ in the direction $p^a/P$. 


\subsection{The empirical distance of two particles}
\label{sub-2.2}

The aim of the present subsection is to express the distance between any 
two straight lines of the Euclidean 3-space, considered to be the 
centre-of-mass lines of elementary classical systems, by $E(3)$-invariant 
classical observables. 

Let $(p^a_{\bi},J^{ab}_{\bi})$, ${\bi}=1,2$, characterize two elementary 
classical mechanical systems, and let us form their formal union with the 
linear and angular momenta $p^a:=p^a_1+p^a_2$ and $J^{ab}:=J^{ab}_1+J^{ab}_2$, 
respectively. Then $P^2$ and $W$ for the composite system are defined in 
terms of $p^a$ and $J^{ab}$ according to the general rules (\ref{eq:2.1.1}). 
Since the Lie algebra of the basic observables of the composite system is 
the direct sum of the Lie algebras of those of the constituent subsystems, 
it is easy to check that $p^a$ and $J^{ab}$ satisfy the commutation relations 
(\ref{eq:2.1.2a})-(\ref{eq:2.1.2c}), i.e. generate the Lie algebra $e(3)$; 
and that $P^2$, $P_1^2$, $P_2^2$ and also $W$, $W_1$, $W_2$ are all commuting 
with $p^a$ and $J^{ab}$. Hence, $P^2_{12}:=\delta_{ab}p^a_1p^b_2=\frac{1}{2}
(P^2-P_1^2-P_2^2)$ and $W_{12}:=W-W_1-W_2$ are also commuting with $p^a$ and 
$J^{ab}$. It is $P^2_{12}$ and $W_{12}$ (and their quantum mechanical version) 
that play fundamental role in the subsequent analysis. 

Since $P_1P_2\not=0$, by (\ref{eq:2.1.1}) and (\ref{eq:2.1.3}) we obtain 
\begin{equation}
W_{12}=\frac{1}{2}J^{ab}_1\varepsilon_{abc}p^c_2+\frac{1}{2}J^{ab}_2
\varepsilon_{abc}p^c_1=W_1\frac{P^2_{12}}{P_1^2}+W_2\frac{P^2_{12}}{P_2^2}+
\bigl(\frac{M^a_1}{P^2_1}-\frac{M^a_2}{P^2_2}\bigr)\varepsilon_{abc}p^b_1p^c_2.
\label{eq:2.2.1}
\end{equation}
Since our aim is to express $M^a_1/P^2_1-M^a_2/P^2_2$ from (\ref{eq:2.2.1}), 
for a moment we assume that the linear momenta of the constituent systems 
are linearly independent. This requirement is equivalent to the condition 
$P_1^2P_2^2>P^4_{12}$. In this case, the last term on the right in 
(\ref{eq:2.2.1}) is not zero, and then this equation can be solved for 
$M^a_1/P^2_1-M^a_2/P^2_2$. This solution is 
\begin{equation}
\frac{M^a_1}{P^2_1}-\frac{M^a_2}{P^2_2}=\frac{1}{P_1^2P_2^2-P^4_{12}}\Bigl(
W_{12}-W_1\frac{P^2_{12}}{P_1^2}-W_2\frac{P^2_{12}}{P_2^2}\Bigr)\varepsilon^a
{}_{bc}p^b_1p^c_2+u_1p^a_1+u_2p^a_2 \label{eq:2.2.2}
\end{equation}
for arbitrary $u_1,u_2\in\mathbb{R}$. Although its components in the 2-plane 
spanned by $p^a_1$ and $p^a_2$ are ambiguous, its component in the direction 
orthogonal to both $p^a_1$ and $p^a_2$, 
\begin{equation}
d_{12}:=\bigl(\frac{M^a_1}{P^2_1}-\frac{M^a_2}{P^2_2}\bigr)\frac{\varepsilon
_{abc}p^b_1p^c_2}{\sqrt{P_1^2P_2^2-P^4_{12}}}=\frac{1}{\sqrt{P_1^2P_2^2-P^4_{12}}}
\Bigl(W_{12}-W_1\frac{P^2_{12}}{P_1^2}-W_2\frac{P^2_{12}}{P_2^2}\Bigr),
\label{eq:2.2.3}
\end{equation}
is well defined. Similarly, although under a translation $M^a_1/P^2_1-M^a_2
/P^2_2$ changes as $(M^a_1/P^2_1-M^a_2/P^2_2)\mapsto(M^a_1/P^2_1-M^a_2/P^2_2)-
(p^a_1p^b_1/P^2_1-p^a_2p^b_2/P^2_2)\xi_b$, its component in the direction 
orthogonal both to $p^a_1$ and $p^a_2$ is invariant. Therefore, the vector 
$d^a_{12}$ defined by the right hand side of (\ref{eq:2.2.2}) with $u_1=u_2
=0$ is uniquely determined, it is orthogonal both to $p^a_1$ and $p^a_2$ 
and points from a uniquely determined point $\nu_{21}$ of the straight line 
$q^a_2(u):=q^a_2+up^a_2$, $u\in\mathbb{R}$, to a uniquely determined point 
$\nu_{12}$ of the straight line $q^a_1(u):=q^a_1+up^a_1$, $u\in\mathbb{R}$. 
Its physical dimension is length, and it is invariant with respect to the 
$P_1\mapsto\alpha P_1$, $P_2\mapsto\beta P_2$ rescalings for any $\alpha,
\beta>0$. Thus $d^a_{12}$ is the \emph{relative position vector} of the 
first subsystem with respect to the second; and $d_{12}$ is the (signed) 
\emph{distance} between the \emph{centre-of-mass lines} of the two 
constituent, elementary subsystems. In particular, $d_{12}=0$ holds 
precisely when the two centre-of-mass lines intersect each other. This can 
always be achieved by an appropriate translation of \emph{one of the two} 
subsystems, and hence $d^a_{12}$ can always be characterized by such a 
translation. 

If $p^a_1$ and $p^a_2$ are parallel, then, as we are going to show, the 
distance of the two corresponding centre-of-mass lines that do not coincide 
can be recovered as the limit of the distances between centre-of-mass lines 
with non-parallel tangents, i.e. as a limit of classical observables above. 
Since both the numerator and the denominator in (\ref{eq:2.2.3}) are 
vanishing if $p^a_1$ and $p^a_2$ are parallel, we should check that these 
tend to zero in the same order and that their quotient is finite and well 
defined. Since the two centre-of-mass lines are parallel but do not coincide, 
they lay in a 2-plane, and the vector $M^a_1/P^2_1-M^a_2/P^2_2$ is non-zero, 
tangent to this 2-plane and is orthogonal to $p^a_1$ (and hence to $p^a_2$, 
too). Thus, there is a unit vector $w^a$ which is orthogonal both to $p^a_1$ 
and $M^a_1/P^2_1-M^a_2/P^2_2$. Then let us consider the 1-parameter family of 
momenta $p^a_2(\alpha):=P_2(\cos\alpha\,v^a+\sin\alpha\,w^a)$, where $v^a:=
p^a_1/P_1=p^a_2/P_2$. In the $\alpha\to0$ limit, $p^a_2(\alpha)\to p^a_2$. A 
straightforward calculation gives that $P_1^2P^2_2(\alpha)-P^4_{12}(\alpha)=
P^2_1P^2_2\sin^2\alpha$, $\varepsilon^a{}_{bc}p^b_1p^c_2(\alpha)=P_1P_2
\varepsilon^a{}_{bc}v^bw^c\sin\alpha$ and 
\begin{eqnarray*}
W_{12}(\alpha)\!\!\!\!&-\!\!\!\!&W_1\frac{P^2_{12}(\alpha)}{P^2_1}-
 W_2(\alpha)\frac{P^2_{12}(\alpha)}{P^2_2}= \\
\!\!\!\!&=\!\!\!\!&\frac{1}{2}\Bigl(J^{ab}_1\varepsilon_{abc}w^cP_2-J^{ab}_2
 \varepsilon_{abc}w^cP_1\cos\alpha+J^{ab}_2\varepsilon_{abc}p^c_1\sin\alpha\Bigr)
 \sin\alpha.
\end{eqnarray*}
Hence, the $\alpha\to0$ limit of $d^a_{12}(\alpha)$ for the centre-of-mass 
lines with the tangents  $p^a_1$ and $p^a_2(\alpha)$ is, indeed, the well
defined finite value 
\begin{equation*}
\frac{1}{2}\Bigl(\frac{J^{de}_1P_2-J^{de}_2P_1}{P_1P_2}\varepsilon_{def}w^f\Bigr)
\varepsilon^a{}_{bc}v^bw^c=\bigl(\frac{M^d_1}{P^2_1}-\frac{M^d_2}{P^2_2}\bigr)
\varepsilon_{def}v^ew^f\,\varepsilon^a{}_{bc}v^bw^c. 
\end{equation*}
Hence, the distance $d_{12}$ is also well defined and finite. In this case, 
however, the points analogous to $\nu_{12}$ and $\nu_{21}$ above are 
undetermined. 

To summarize: we found an alternative expression $\vert d_{\bi\bj}\vert$ for 
the Euclidean distance, 
\begin{equation}
D_{\bi\bj}:=\inf\Bigl\{\sqrt{\delta_{ab}(q^a_{\bi}(u_{\bi})-q^a_{\bj}(u_{\bj}))
(q^b_{\bi}(u_{\bi})-q^b_{\bj}(u_{\bj}))}\,\vert\, u_{\bi},u_{\bj}\in\mathbb{R}
\Bigr\}, \label{eq:2.2.4}
\end{equation}
between the centre-of-mass straight lines of any two elementary classical 
systems \emph{in terms of $E(3)$-invariant basic observables of the composite 
system}. Since the latter is given by the \emph{observables} of the 
composite system, we call $d_{\bi\bj}$ the `empirical distance' between the 
two subsystems. However, if we know the distance between any two straight 
lines in $\mathbb{R}^3$, then we can determine the Euclidean distance 
function, too. Hence, our empirical distance determines the metric structure 
of $\mathbb{R}^3$. We will see that the analogous empirical distance in 
quantum theory deviates from the \emph{a priori} Euclidean distance 
$D_{\bi\bj}$, but in the classical limit the former reproduces the latter. 


\subsection{Empirical angles and volume elements}
\label{sub-2.3}

In \cite{Sz21c}, we considered $SO(3)$-invariant elementary classical 
systems in which the only basic observable was the angular momentum vector 
$J^a$; and we introduced an empirical angle between the angular momenta of 
two subsystems in a well defined and $SO(3)$-invariant way. However, in the 
present case the symmetry group is the Euclidean group $E(3)$, and although 
the analogously defined empirical angles would still be $SO(3)$-invariant, 
but they would \emph{not} be invariant under translations. Thus, in 
$E(3)$-invariant classical systems, the empirical angles between the 
subsystems should be defined in terms of quantities that are invariant under 
translations and covariant under rotations. 

The primary candidate for such quantities is the linear momentum. Thus we 
define the `empirical angle' $\omega_{12}$ between $p^a_1$ and $p^a_2$ 
according to 
\begin{equation}
\cos\omega_{12}:=\frac{\delta_{ab}p^a_1p^b_2}{P_1P_2} \label{eq:2.3.1}
\end{equation}
with range $\omega_{12}\in[0,\pi]$. 
This angle is clearly $E(3)$-invariant, and if e.g. $p^a_1/P_1$ can be 
obtained from $p^a_2/P_2$ by a rotation  with angle $\beta_{12}$ in the 
2-plane spanned by $p^a_1$ and $p^a_2$, then $\omega_{12}=\beta_{12}$. Thus 
$\omega_{12}$ reproduces the angles of the \emph{a priori} Euclidean geometry 
of $\mathbb{R}^3$ (see also \cite{Sz21c}). However, as we will see, they 
split in the quantum theory, and they coincide only in the classical limit. 

The natural volume 3-form $\varepsilon_{abc}$ on the space of the translation 
generators in $e(3)$ makes it possible to introduce the `empirical 3-volume 
element' by the 3-volume of the tetrahedron spanned by three linear momenta, 
$p^a_1$, $p^a_2$ and $p^a_3$, of a three-particle system by 
\begin{equation}
v_{123}:=\frac{1}{3!}\varepsilon_{abc}\frac{p^a_1p^b_2p^c_3}{P_1P_2P_2}. 
\label{eq:2.3.2}
\end{equation}
Just in the case of empirical angles, in quantum theory the 
corresponding empirical 3-volume element and the 3-volume element of the 
\emph{a priori} Euclidean geometry of $\mathbb{R}^3$ do not coincide. 
They do only in the classical limit (see also \cite{Sz21c}). In Section 
\ref{sec-5} we raise the possibility of another notion of empirical angles 
and 3-volume elements, based on the relative position vectors $d^a_{\bi\bj}$ 
of three-particle systems, rather than the linear momenta. 


\section{$E(3)$-invariant elementary quantum mechanical systems}
\label{sec-3}

\subsection{The definition and the basic properties of the elementary 
systems}
\label{sub-3.1}

Adapting the idea of Poincar\'e-invariant \emph{elementary} quantum 
mechanical systems of Newton and Wigner \cite{NeWign} to the present case, 
a \emph{Euclidean-invariant elementary quantum mechanical system} will be 
defined to be a system whose states belong to the carrier space of some 
\emph{unitary, irreducible} representation of the (quantum mechanical) 
Euclidean group $E(3)$; and, in this representation, the momentum and 
angular momentum tensor operators, ${\bf p}^a$ and ${\bf J}^{ab}$, are the 
self-adjoint generators of the translations and rotations, respectively. 
These representations are labelled by a fixed value $P^2\geq0$ and $w$, 
respectively, of the two Casimir operators 
\begin{equation}
{\bf P}^2:=\delta_{ab}{\bf p}^a{\bf p}^b, \hskip 20pt
{\bf W}:=\frac{1}{2}\varepsilon_{abc}{\bf J}^{ab}{\bf p}^c. \label{eq:3.1.1}
\end{equation}
Thus, the quantum mechanical operators will be denoted by boldface letters. 
Clearly, in the states belonging to such representations, the energy ${\bf 
E}^2={\bf P}^2+m^2$ or ${\bf E}={\bf P}^2/2m$ of the relativistic or 
non-relativistic systems of rest mass $m$, respectively, has a definite 
value. The commutators of ${\bf p}^a$ and ${\bf J}^{cd}$ are 
\begin{eqnarray}
&{}&[{\bf p}^a,{\bf p}^b]=0, \hskip 20pt
 [{\bf p}^a,{\bf J}^{cd}]=-{\rm i}\hbar\bigl(\delta^{ac}{\bf p}^d-\delta^{ad}
 {\bf p}^c\bigr), \label{eq:3.1.2a} \\
&{}&[{\bf J}^{ab},{\bf J}^{cd}]=-{\rm i}\hbar\bigl(\delta^{bc}{\bf J}^{ad}-
 \delta^{bd}{\bf J}^{ac}+\delta^{ad}{\bf J}^{bc}-\delta^{ac}{\bf J}^{bd}\bigr);
 \label{eq:3.1.2b}
\end{eqnarray}
which are just the Lie brackets (\ref{eq:2.1.2a})-(\ref{eq:2.1.2c}) with the 
$p^a\mapsto{\bf p}^a$ and $J^{ab}\mapsto({\rm i}/\hbar){\bf J}^{ab}$ 
substitution. 

Nevertheless, ${\bf M}_a:={\bf J}_{ab}{\bf p}^b$ is \emph{not} self-adjoint, 
because ${\bf M}_a^\dagger={\bf p}^b{\bf J}_{ab}={\bf M}_a+[{\bf p}^b,{\bf J}
_{ab}]={\bf M}_a+2{\rm i}\hbar{\bf p}_a$. Thus we form ${\bf C}_a:=\frac{1}{2}
({\bf M}_a+{\bf M}^\dagger_a)={\bf M}_a+{\rm i}\hbar{\bf p}_a$, which is, by 
definition, the self-adjoint part of ${\bf M}_a$, and we consider this to be 
the centre-of-mass operator. The commutators of these operators can be 
derived from those for ${\bf p}^a$ and ${\bf J}^{ab}$ above: 
\begin{equation}
[{\bf p}_a,{\bf C}_b]=-{\rm i}\hbar(\delta_{ab}{\bf P}^2-{\bf p}_a{\bf p}_b), 
\hskip 20pt
[{\bf C}_a,{\bf C}_b]=-{\rm i}\hbar {\bf P}^2{\bf J}_{ab}. \label{eq:3.1.3}
\end{equation}
As a consequence of the definitions, 
\begin{equation}
{\bf P}^2{\bf J}_{ab}={\bf C}_a{\bf p}_b-{\bf C}_b{\bf p}_a+\varepsilon_{abc}
{\bf p}^c{\bf W}, \label{eq:3.1.4}
\end{equation}
i.e. the analog of (\ref{eq:2.1.3}) holds for the operators, too. 

The unitary, irreducible representations of the quantum mechanical Euclidean 
group has been determined in various different forms (see e.g. 
\cite{Rno,Tung,KoRe,SiJa}). The form that we use in the present paper is 
given in the appendices of \cite{Sz21a,Sz21b}, and is based on the use of 
the complex line bundles ${\cal O}(-2s)$ over the 2-sphere ${\cal S}_P$ of 
radius $P$ in the classical momentum 3-space, where $2s\in\mathbb{Z}$. Here 
${\cal O}(-2s)$ is the bundle of spin weighted scalars on ${\cal S}_P$ with 
spin weight $s$ (see e.g. \cite{HT}). The wave functions of the quantum 
system are square integrable cross sections $\phi$ of ${\cal O}(-2s)$, and 
the Hilbert space of these cross sections will be denoted by ${\cal H}_{P,s}$, 
or simply by ${\cal H}$. Thus, ${\cal S}_P$ is analogous to the mass shell 
of the Poincar\'e invariant systems (see e.g. \cite{StWi}). The radius $P$ 
is fixed by the Casimir operator ${\bf P}^2$. The spin weight $s$ of $\phi$ 
is linked to the value of the other Casimir operator, ${\bf W}$, in the 
irreducible representation: $w=\hbar Ps$. 

The $SU(2)$ part of $E(3)$ acts on ${\cal S}_P$ as $p^a\mapsto R^a{}_bp^b$, 
where $R^a{}_b$ is the rotation matrix determined by $U^A{}_B\in SU(2)$ via 
$R^a{}_b:=-\sigma^a_{AA'}U^A{}_B\bar U^{A'}{}_{B'}\sigma^{BB'}_b$. Here $\sigma_a
^{AA'}$ are the three non-trivial $SL(2,\mathbb{C})$ Pauli matrices (including 
the factor $1/\sqrt{2}$), according to the conventions of \cite{PR1}. 
Raising/lowering of the unprimed and primed spinor name indices, $A,B$ and 
$A',B'$, are defined by the spinor metric $\varepsilon_{AB}$ and its complex 
conjugate, 
respectively. Then the action of $SU(2)$ on the spin weighted function 
$\phi$ is $\phi(p^e)\mapsto\exp(-2{\rm i}s\lambda)\phi((R^{-1})^e{}_cp^c)$; 
while that of the translation by $\xi^a$ is simply\footnote{In \cite{Sz21b} 
we defined the translation by multiplication by the phase factor $\exp(
{\rm i}p_a\xi^a/\hbar)$. Here, to be compatible with the standard sign
convention, we changed the sign in the exponent to its opposite.} $\phi(p^e)
\mapsto\exp(-{\rm i}p_a\xi^a/\hbar)\phi(p^e)$. Here, $\exp({\rm i}\lambda)$ 
is just the phase that appears in the action, $\pi^A\mapsto\exp({\rm i}
\lambda)\pi^A$, of $SU(2)$ on the spinor constituent $\pi^A$ of $p^a$ (see 
\cite{Sz21a,Sz21b}). This representation of $E(3)$ is analogous to that of 
the (quantum mechanical) Poincar\'e group on the $L_2$-space of spinor 
fields on the mass shell \cite{StWi}. 

The spin weighted spherical harmonics ${}_sY_{jm}$ with spin weight $s$ and 
indices $j=\vert s\vert,\vert s\vert+1,...$, $m=-j,-j+1,...,j$ are known to 
form an orthonormal basis in the space of the square integrable cross 
sections of ${\cal O}(-2s)$ on the \emph{unit} 2-sphere (see \cite{PR1,NP}). 
Hence, the space that the functions ${}_sY_{jm}$ on ${\cal S}_R$ span is 
\emph{precisely} the carrier space of the unitary, irreducible 
representation of $E(3)$ labelled by the Casimir invariants $P>0$ and $w=
\hbar Ps$, in which ${}_sY_{jm}/P$ form an \emph{orthonormal} basis with 
respect to the natural $L_2$ scalar product. 

In this representation, the action of ${\bf p}_a$ and ${\bf J}_a:=\frac{1}{2}
\varepsilon_{abc}{\bf J}^{bc}$ on $\phi$ is 
\begin{eqnarray}
&{}&{\bf p}_a\phi=p_a\phi, \label{eq:3.1.5a} \\
&{}&{\bf J}_a\phi=P\hbar\Bigl(m_a{\edth}'\phi-\bar m_a{\edth}\phi\Bigr)+
  s\hbar\frac{p_a}{P}\phi, \label{eq:3.1.5b}
\end{eqnarray}
where $m^a$ and $\bar m^a$ are the complex null tangents of ${\cal S}_P$ and 
normalized by $m_a\bar m^a=1$, and ${\edth}$ and ${\edth}'$ are the edth 
operators of Newman and Penrose \cite{NP}. The explicit form of the vectors 
$p^a$, $m^a$ and $\bar m^a$ as well as the operators ${\edth}$ and ${\edth}'$ 
in the complex stereographic coordinates $(\zeta,\bar\zeta)$ on ${\cal S}_P$ 
is given in Appendix \ref{sub-A.1}. The first two terms together on the 
right of (\ref{eq:3.1.5b}) give the orbital part of the angular momentum 
operator, denoted by ${\bf L}_a$, while the third its spin part. 
Since the spin weighted harmonics form a basis in ${\cal H}_{P,s}$, the basic 
observables ${\bf p}_a$ and ${\bf J}_a$ can also be given by their action on 
these harmonics, too. Their matrix elements are calculated in Appendix 
\ref{sub-A.2}. 

${\bf p}^a$ and ${\bf J}^a$ are known to be $SO(3)$ vector operators. Thus, 
if the unitary operator ${\bf U}$ represents $U^A{}_B\in SU(2)$ on ${\cal H}
_{P,s}$, then 
\begin{equation}
{\bf U}^\dagger{\bf p}^a{\bf U}=R^a{}_b{\bf p}^b, \hskip 20pt
{\bf U}^\dagger{\bf J}^a{\bf U}=R^a{}_b{\bf J}^b. \label{eq:3.1.6}
\end{equation}
If $\phi=\exp(-{\rm i}p_a\xi^a/\hbar)\chi$, then by (\ref{eq:3.1.5a}) and 
(\ref{eq:3.1.5b}) 
\begin{equation}
{\bf p}^a\phi=\exp\bigl(-\frac{\rm i}{\hbar}p_e\xi^e\bigr){\bf p}^a\chi,
\hskip 20pt
{\bf J}^a\phi=\exp\bigl(-\frac{\rm i}{\hbar}p_e\xi^e\bigr)\Bigl({\bf J}^a
  \chi+\varepsilon^a{}_{bc}\xi^b{\bf p}^c\chi\Bigr). \label{eq:3.1.6a}
\end{equation}
Hence, for $\phi=\exp(-{\rm i}p_a\xi^a/\hbar){\bf U}\psi$, 
\begin{equation}
\langle\phi\vert{\bf p}^a\vert\phi\rangle=R^a{}_b\langle\psi\vert{\bf p}^b
  \vert\psi\rangle, \hskip 20pt
\langle\phi\vert{\bf J}^a\vert\phi\rangle=R^a{}_b\langle\psi\vert{\bf J}^b
  \vert\psi\rangle+\varepsilon^a{}_{bc}\xi^bR^c{}_d\langle\psi\vert{\bf p}^d
  \vert\psi\rangle. \label{eq:3.1.7}
\end{equation}
These are just the transformation laws of the classical basic observables 
under the action of the classical $E(3)$. 

Finally, it is straightforward to derive the explicit form of the 
centre-of-mass operator: 
\begin{equation}
{\bf C}^a\phi={\rm i}\hbar\Bigl(P^2m^a{\edth}'\phi+P^2\bar m^a{\edth}\phi-
p^a\phi\Bigr). \label{eq:3.1.8}
\end{equation}
This is also an $SO(3)$ vector operator. A detailed discussion of the line 
bundle ${\cal O}(-2s)$ and the derivation of these equations are also given 
in the appendices of \cite{Sz21a,Sz21b}. 


\subsection{The centre-of-mass states}
\label{sub-3.2}

In this subsection we show that the spin weighted spherical harmonics 
form a distinguished basis among the orthonormal bases in ${\cal H}_{P,s}$ 
in the sense that \emph{they are adapted in a natural way to the basis in 
the abstract Lie algebra $e(3)$ of the basic quantum observables} (but 
\emph{not} to a Cartesian frame in the `physical 3-space'). In particular, 
they are just the eigenfunctions of the square of the centre-of-mass vector 
operator, ${\bf C}_a{\bf C}^a$, they are the critical points of the functional 
$\phi\mapsto\langle\phi\vert{\bf C}_a{\bf C}^a\vert\phi\rangle$, and the 
expectation value of the centre-of-mass vector operator in these states is 
zero. We will calculate the expectation value of our empirical distance and 
angle in the states that are obtained from these special ones by some $E(3)$ 
transformation. 

Using (\ref{eq:3.1.8}) and ${\edth}p^a=m^a$ (see Appendix \ref{sub-A.1}), by 
integration by parts it is straightforward to form the centre-of-mass-square 
operator ${\bf C}_a{\bf C}^a$. It is 
\begin{equation}
{\bf C}_a{\bf C}^a\phi=\hbar^2P^2\Bigl(-P^2\bigl({\edth}{\edth}'+{\edth}'
{\edth}\bigr)\phi+\phi\Bigr). \label{eq:3.2.1}
\end{equation}
Since the operators ${\bf C}_a$ are self-adjoint, ${\bf C}_a{\bf C}^a$ is a 
positive self-adjoint operator. Similar calculations yield the square of 
the total and the orbital angular momentum operators: 
\begin{equation*}
{\bf J}_a{\bf J}^a\phi=\hbar^2\Bigl(-P^2\bigl({\edth}{\edth}'+{\edth}'{\edth}
\bigr)\phi+s^2\phi\Bigr), \hskip 20pt
{\bf L}_a{\bf L}^a\phi=-P^2\hbar^2\bigl({\edth}{\edth}'+{\edth}'{\edth}
\bigr)\phi. 
\end{equation*}
Thus ${\bf C}_a{\bf C}^a$, ${\bf J}_a{\bf J}^a$ and ${\bf L}_a{\bf L}^a$ 
deviate from one another only by a constant times the identity operator, and 
hence, in particular, their spectral properties are the same. Note also that 
$({\edth}{\edth}'+{\edth}'{\edth})$ is just the metric Laplace operator on 
${\cal S}_P$. 

First we show that the critical points of the functional $\phi\mapsto\langle
\phi\vert{\bf C}_a{\bf C}^a\vert\phi\rangle$ on ${\cal H}_{P,s}$ are just the 
combinations of the form $\phi=\sum_mc^m{}_sY_{jm}$ of spin weighted spherical 
harmonics ${}_sY_{jm}$ with given $j$. Using $\langle\phi\vert\phi\rangle=1$, 
by integration by parts (\ref{eq:3.2.1}) gives 
\begin{equation*}
\langle\phi\vert{\bf C}_a{\bf C}^a\vert\phi\rangle=\hbar^2P^2\Bigl(P^2\int
_{{\cal S}_P}\bigl(({\edth}\bar\phi)({\edth}'\phi)+({\edth}'\bar\phi)
({\edth}\phi)\bigr){\rm d}{\cal S}_P+1\Bigr). 
\end{equation*}
The critical points of this functional with respect to the variations 
$\delta\phi$ constrained by $\langle\phi\vert\phi\rangle=1$ are just the 
critical points of the functional $\langle\phi\vert{\bf C}_a{\bf C}^a\vert
\phi\rangle+\lambda\hbar^2P^4\langle\phi\vert\phi\rangle$ with respect to 
unconstrained variations, where $\lambda$ is some real Lagrange multiplier. 
The vanishing of the variation of the latter functional yields 
\begin{equation}
\bigl({\edth}{\edth}'+{\edth}'{\edth}\bigr)\phi=\lambda\phi. \label{eq:3.2.2}
\end{equation}
Thus $\lambda$ must be an eigenvalue of the Laplace operator ${\edth}{\edth}'
+{\edth}'{\edth}$ acting on spin weight $s$ functions on ${\cal S}_P$, and 
then the critical configurations are given by the corresponding 
eigenfunctions. These can be determined by expanding $\phi$ as $\phi=\sum
_{j,m}c^{jm}{}_sY_{jm}$ with complex constants $c^{jm}$, where $j=\vert s\vert,
\vert s\vert+1,\vert s\vert+2,...$ and $m=-j,-j+1,...,j$; and using the 
general formulae (\ref{eq:A.1.7}) how the operators ${\edth}$ and ${\edth}'$ 
act on ${}_sY_{jm}$. Substituting all these into (\ref{eq:3.2.2}), we find 
that $\lambda=-P^{-2}(j^2+j-s^2)$. Hence, $\lambda$ is linked to $j$, and the 
corresponding eigenfunctions have the form $\phi=\sum_mc^m\,{}_sY_{jm}$. 
These eigenfunctions are the critical points of the functional $\langle\phi
\vert{\bf C}_a{\bf C}^a\vert\phi\rangle$, and the corresponding critical 
values are 
\begin{equation}
\langle\phi\vert{\bf C}_a{\bf C}^a\vert\phi\rangle=\hbar^2P^2\bigl(1+j^2+j-
s^2\bigr)\geq\hbar^2P^2\bigl(1+\vert s\vert\bigr). \label{eq:3.2.3}
\end{equation}
The smallest of these corresponds to $j=\vert s\vert$, and the corresponding 
eigenfunction is $\phi=\sum_mc^m\,{}_sY_{\vert s\vert m}$, which is 
\emph{holomorphic} if $s=-\vert s\vert$, and it is \emph{anti-holomorphic} 
if $s=\vert s\vert$ (see Appendix \ref{sub-A.1}). Forming the second 
variation of $\langle\phi\vert{\bf C}_a{\bf C}^a\vert\phi\rangle+\lambda
\hbar^2P^4\langle\phi\vert\phi\rangle$ at the critical points, one can see 
that the only minimum does, in fact, correspond to $j=\vert s\vert$, and 
all the other critical points are only inflection. Thus the right hand 
side of (\ref{eq:3.2.3}) is the sharp \emph{strictly positive} lower bound 
for the expectation values of ${\bf C}_a{\bf C}^a$. 

This analysis shows also that the spectrum of all the operators ${\bf C}_a
{\bf C}^a$, ${\bf J}_a{\bf J}^a$ and ${\bf L}_a{\bf L}^a$ is discrete (as it 
must be since ${\edth}{\edth}'+{\edth}'{\edth}$ is an \emph{elliptic} 
differential operator acting on cross sections of a vector bundle over a 
\emph{compact} manifold), their eigenvalues, respectively, are 
\begin{equation*}
P^2\hbar^2(1+j^2+j-s^2), \hskip 20pt
\hbar^2\,j(j+1), \hskip 20pt
\hbar^2(j^2+j-s^2);
\end{equation*}
and the corresponding common eigenfunctions are of the form $\sum_mc^m{}_s
Y_{jm}$. These imply, in particular, that the expectation value of any of 
these operators is not zero for $s\not=0$. 

By integration by parts and using how the edth operators act on $p_a$, $m_a$ 
and $\bar m_a$, we can write 
\begin{eqnarray*}
\langle\phi\vert{\bf C}_a\vert\phi\rangle\!\!\!\!&=\!\!\!\!&{\rm i}\hbar\,
 P^2\int_{{\cal S}_P}\bar\phi\bigl(m_a\,{\edth}'\phi+\bar m_a\,{\edth}\phi
 +\phi\,{\edth}\bar m_a\bigr){\rm d}{\cal S}_P=\\
\!\!\!\!&=\!\!\!\!&{\rm i}\hbar\,P^2\int_{{\cal S}_p}\bigl(\bar\phi\, m_a\,
 {\edth}'\phi-\phi\,\bar m_a\,{\edth}\bar\phi\bigr){\rm d}{\cal S}_P=
 {\rm i}\hbar\,P^2\int_{{\cal S}_P}p_a\bigl(\phi\,{\edth}'{\edth}\bar\phi-
 \bar\phi\,{\edth}{\edth}'\phi\bigr){\rm d}{\cal S}_P.
\end{eqnarray*}
Since $\phi=\sum_mc^m\,{}_sY_{jm}$ for some given $j$, by (\ref{eq:A.1.7}) 
we have that $\bar\phi({\edth}{\edth}'\phi)=-(j+s)(j-s+1)\bar\phi\phi/2P^2$. 
Using this and its complex conjugate, finally we obtain 
\begin{equation}
\langle\phi\vert{\bf C}_a\vert\phi\rangle=0. \label{eq:3.2.4}
\end{equation}
Thus, the $2j+1$ dimensional subspaces in ${\cal H}_{P,s}$ spanned by the 
harmonics ${}_sY_{jm}$ with given $j$ are specified e.g. by ${\bf C}_a{\bf C}
^a$ (or, equivalently by ${\bf J}_a{\bf J}^a$ or by ${\bf L}_a{\bf L}^a$) in a 
natural way, while the basis in these subspaces, the index $m$ is referring 
to, is linked to our choice for the basis \emph{in the sub-Lie algebra 
$su(2)\subset e(3)$}. 

To summarize, $\{{}_sY_{jm}/P\}$ is not only one of the many $L_2$-orthonormal 
bases in ${\cal H}_{P,s}$, but it is \emph{adapted in a natural way to the 
centre-of-mass operator}, too. Although the expectation value of ${\bf C}_a$ 
is zero in any eigenstate of ${\bf C}_a{\bf C}^a$, the expectation value of 
${\bf C}_a{\bf C}^a$ can \emph{never} be zero, even if $s=0$. Its smallest 
expectation value, which is its smallest eigenvalue, cannot be made zero 
e.g. by any translation (in contrast to the classical case). It corresponds 
to $j=\vert s\vert$, and the corresponding eigenfunctions, $\phi=\sum_mc^m
{}_sY_{\vert s\vert m}$, are \emph{holomorphic} if $s=-\vert s\vert$, and 
\emph{anti-holomorphic} if $s=\vert s\vert$. We call these states the 
\emph{centre-of-mass states}. These form a $2\vert s\vert+1$ dimensional 
subspace in ${\cal H}_{P,s}$, and these are the states of the 
$E(3)$-invariant elementary quantum mechanical systems that are the closest 
analogs of the states of the classical systems with vanishing centre-of-mass 
vector. 


\section{The two-particle system}
\label{sec-4}

\subsection{The quantum observables of two-particle systems}
\label{sub-4.1}

Let us consider two $E(3)$-invariant elementary quantum mechanical systems, 
whose basic quantum observables are ${\bf p}^a_{\bi}$ and ${\bf J}^{ab}_{\bi}$, 
${\bi}=1,2$. These observables are self-adjoint operators on ${\cal H}_{\bi}$. 
The corresponding Casimir operators are denoted by ${\bf P}^2_{\bi}$ and 
${\bf W}_{\bi}$. The Hilbert space of the joint system is ${\cal H}_1\otimes
{\cal H}_2$, and we can form the operators ${\bf O}_1\otimes{\bf I}_2$, 
${\bf I}_1\otimes{\bf O}_2:{\cal H}_1\otimes{\cal H}_2\to{\cal H}_1\otimes
{\cal H}_2$ for any ${\bf O}_{\bi}:{\cal H}_{\bi}\to{\cal H}_{\bi}$, where 
${\bf I}_{\bi}$ are the identity operators on the respective Hilbert spaces 
${\cal H}_{\bi}$. Clearly, ${\bf O}_1\otimes{\bf I}_2$ and ${\bf I}_1\otimes
{\bf O}_2$ are commuting. In particular, ${\bf P}^2_1\otimes{\bf I}_2$, 
${\bf W}_1\otimes{\bf I}_2$, ${\bf I}_1\otimes{\bf P}^2_2$ and ${\bf I}_1
\otimes{\bf W}_2$ are Casimir operators of the composite system. 

Analogously to (\ref{eq:3.1.1}), we form 
\begin{eqnarray}
{\bf P}^2:=\!\!\!\!&{}\!\!\!\!&\delta_{ab}\bigl({\bf p}^a_1\otimes{\bf I}_2+
  {\bf I}_1\otimes{\bf p}^a_2\bigr)\bigl({\bf p}^b_1\otimes{\bf I}_2+{\bf I}_1
  \otimes{\bf p}^b_2\bigr)= \nonumber \\
=\!\!\!\!&{}\!\!\!\!&{\bf P}^2_1\otimes{\bf I}_2+{\bf I}_1\otimes{\bf P}^2_2
  +2\delta_{ab}{\bf p}^a_1\otimes{\bf p}^b_2, \label{eq:4.1.1a} \\
{\bf W}:=\!\!\!\!&{}\!\!\!\!&\frac{1}{2}\varepsilon_{abc}\bigl({\bf J}^{ab}_1
  \otimes{\bf I}_2+{\bf I}_1\otimes{\bf J}^{ab}_2\bigr)\bigl({\bf p}^c_1
  \otimes{\bf I}_2+{\bf I}_1\otimes{\bf p}^c_2\bigr)= \nonumber \\
=\!\!\!\!&{}\!\!\!\!&{\bf W}_1\otimes{\bf I}_2+{\bf I}_1\otimes{\bf W}_2+
  \frac{1}{2}\varepsilon_{abc}\bigl({\bf p}^c_1\otimes{\bf J}^{ab}_2+{\bf J}
  ^{ab}_1\otimes{\bf p}^c_2\bigr). \label{eq:4.1.1b}
\end{eqnarray}
Although ${\bf W}$ does \emph{not} commute with any of ${\bf p}^a_1\otimes
{\bf I}_2$, ${\bf I}_1\otimes{\bf p}^a_2$, ${\bf J}^{ab}_1\otimes{\bf I}_2$ 
and ${\bf I}_1\otimes{\bf J}^{ab}_2$, the observables of the \emph{subsystems} 
in the algebra of observables of the composite system, and ${\bf P}^2$ does 
\emph{not} commute with ${\bf J}^{ab}_1\otimes{\bf I}_2$ and ${\bf I}_1
\otimes{\bf J}^{ab}_2$, but both ${\bf W}$ and ${\bf P}^2$ \emph{do} commute 
with ${\bf p}^a_1\otimes{\bf I}_2+{\bf I}_1\otimes{\bf p}^a_2$ and ${\bf J}
^{ab}_1\otimes{\bf I}_2+{\bf I}_1\otimes{\bf J}^{ab}_2$, i.e. with the linear 
and angular momentum operators of the \emph{composite system}. Therefore, 
though ${\bf P}^2$ and ${\bf W}$ are \emph{not} Casimir operators of the 
composite system, they \emph{are} commuting with the generators of the 
symmetry group $E(3)$, i.e. they are $E(3)$-invariant. Moreover, $[{\bf P}
^2,{\bf W}]=0$ also holds. Hence, 
\begin{eqnarray}
&{}&{\bf P}^2_{12}:=\frac{1}{2}\Bigl({\bf P}^2-{\bf P}^2_1\otimes{\bf I}_2-
  {\bf I}_1\otimes{\bf P}^2_2\Bigr)=\delta_{ab}{\bf p}^a_1\otimes{\bf p}^b_2,
  \label{eq:4.1.2a} \\
&{}&{\bf W}_{12}:={\bf W}-{\bf W}_1\otimes{\bf I}_2-{\bf I}_1\otimes{\bf W}_2
  =\frac{1}{2}\varepsilon_{abc}\bigl({\bf p}^c_1\otimes{\bf J}^{ab}_2+{\bf J}
  ^{ab}_1\otimes{\bf p}^c_2\bigr) \label{eq:4.1.2b}
\end{eqnarray}
are also $E(3)$-invariant and $[{\bf P}^2_{12},{\bf W}_{12}]=0$ holds. These 
operators characterize the relationship between the two subsystems in the 
composite system, and hence they will have particular significance for us. 

Using the definitions above and the first of (\ref{eq:3.1.3}), the identity 
(\ref{eq:3.1.4}) yields 
\begin{equation}
\bigl({\bf P}^2_1\otimes{\bf P}^2_2\bigr){\bf W}_{12}=\varepsilon_{abc}\bigl(
{\bf p}^b_1\otimes{\bf p}^c_2\bigr)\bigl({\bf C}^a_1\otimes{\bf P}^2_2-{\bf P}
^2_1\otimes{\bf C}^a_2\bigr)+{\bf P}^2_{12}\bigl({\bf P}^2_1\otimes{\bf W}_2+
{\bf W}_1\otimes{\bf P}^2_2\bigr). \label{eq:4.1.3}
\end{equation}
Let the two subsystems be elementary, characterized by the Casimir invariants
$(P_1,s_1)$ and $(P_2,s_2)$, respectively. Then by (\ref{eq:3.1.5a}) 
${\bf P}^2_{12}$ is a multiplication operator on ${\cal H}_1\otimes{\cal H}_2$, 
and hence, for any $\phi_1\otimes\phi_2\in{\cal H}_1\otimes{\cal H}_2$, 
(\ref{eq:4.1.3}) gives 
\begin{equation}
\varepsilon_{abc}p^b_1p^c_2\Bigl(\frac{{\bf C}^a_1}{P^2_1}\otimes{\bf I}_2-
{\bf I}_1\otimes\frac{{\bf C}^a_2}{P^2_2}\Bigr)\,\phi_1\otimes\phi_2=\Bigl(
{\bf W}_{12}-\hbar\bigl(\frac{s_1}{P_1}+\frac{s_2}{P_2}\bigr){\bf P}^2_{12}
\Bigr)\phi_1\otimes\phi_2. \label{eq:4.1.4}
\end{equation}
This is analogous to the classical equation (\ref{eq:2.2.1}), and the 
operators on both sides are $E(3)$-invariant. Nevertheless, their physical
dimension is momentum times angular momentum, rather than length. Thus, 
just as in the classical case (and motivated by (\ref{eq:2.2.3})), we 
should consider the \emph{component of} ${\bf C}^a_1\otimes{\bf I}_2/P^2_1-
{\bf I}_1\otimes{\bf C}^a_2/P^2_2$ \emph{in the direction} $\varepsilon^a{}
_{bc}p^b_1p^c_2$. 

This is just (\ref{eq:4.1.4}) divided by $\sqrt{P^2_1P^2_2-(\delta_{ab}p^a_1
p^b_2)^2}$, the length of $\varepsilon_{abc}p^b_1p^c_2$, and we could consider
the operator 
\begin{equation}
\frac{\varepsilon_{abc}p^b_1p^c_2}{\sqrt{P^2_1P^2_2-(\delta_{de}p^d_1p^e_2)^2}}
\Bigl(\frac{1}{P^2_1}{\bf C}^a_1\otimes{\bf I}_2-\frac{1}{P^2_2}{\bf I}_1
\otimes{\bf C}^a_2\Bigr). \label{eq:4.1.5}
\end{equation}
This is a well defined, self-adjoint and $E(3)$-invariant operator, which is 
analogous to the classical expression (\ref{eq:2.2.3}). However, in contrast 
to the classical case, the coefficient under the square root sign in the 
denominator is \emph{not} constant. Hence it could be difficult to use this 
expression e.g. in the calculation of the expectation values. To cure this 
difficulty, using 
\begin{equation*}
\Bigl(\frac{{\rm d}^k}{{\rm d}x^k}\frac{1}{\sqrt{1-x}}\Bigr)(0)=\frac{1}{2}\,
\frac{3}{2}\,\frac{5}{2}\cdots\frac{2k-1}{2}=\frac{1}{2^k}\frac{1\cdot2\cdot
3\cdot4\cdots(2k-1)\cdot(2k)}{2\cdot4\cdots(2k)}=\frac{1}{2^{2k}}\frac{(2k)!}
{k!}, 
\end{equation*}
for the Taylor expansion of $1/\sqrt{P^2_1P^2_2-(\delta_{ab}p^a_1p^b_2)^2}$ we 
obtain 
\begin{equation}
\frac{1}{\sqrt{P^2_1P^2_2-(\delta_{ab}p^a_1p^b_2)^2}}=\frac{1}{P_1P_2}\sum_{k=0}
^\infty\frac{1}{2^{2k}}\frac{(2k)!}{(k!)^2}\bigl(\delta_{ab}\frac{p^a_1}{P_1}
\frac{p^b_2}{P_2}\bigr)^{2k}. \label{eq:4.1.6}
\end{equation}
Using this and recalling that ${\bf P}^2_{12}$ is a multiplication operator, 
the above candidate for the `distance operator' becomes an expression of the 
\emph{positive powers} of ${\bf W}_{12}$ and ${\bf P}^2_{12}$. Nevertheless, 
now the infinite series makes the application of the resulting expression 
difficult in practice. Thus, although in principle this might yield a well 
defined \emph{operator} for the distance of the two subsystems, and certainly 
it would be worth studying this, in the present paper we choose a different 
strategy and look for only the `empirical distance'. This is the one that we 
followed in \cite{Sz21c} in defining the angle between the angular momentum 
vectors of $SU(2)$-invariant elementary quantum mechanical systems. 


\subsection{The empirical distance}
\label{sub-4.2}

Based on equation (\ref{eq:4.1.4}) and the discussion above, we define the 
\emph{empirical distance of the two $E(3)$-invariant elementary quantum 
mechanical systems} (characterized by their Casimir invariants $(P_1,s_1)$ 
and $(P_2,s_2)$) in their states $\phi_1$ and $\phi_2$, respectively, by 
\begin{equation}
d_{12}:=\frac{\langle\phi_1\otimes\phi_2\vert{\bf W}_{12}-\hbar\bigl(s_1/{P_1}
+{s_2}/{P_2}\bigr){\bf P}^2_{12}\vert\phi_1\otimes\phi_2\rangle}{\sqrt{P^2_1
P^2_2-\langle\phi_1\otimes\phi_2\vert{\bf P}^4_{12}\vert\phi_1\otimes\phi_2
\rangle}}. \label{eq:4.2.1}
\end{equation}
$d_{\bi\bj}$ can, in fact, be defined in any state of the composite system 
consisting of any number of elementary systems, ${\bi},{\bj}=1,\cdots,N$, 
represented by a general density operator $\rho:{\cal H}_{P_1,s_1}\otimes
\cdots\otimes{\cal H}_{P_N,s_N}\to{\cal H}_{P_1,s_1}\otimes\cdots\otimes{\cal H}
_{P_N,s_N}$, not only in pure tensor product states of a bipartite system. 
However, if the density operator represents a pure vector state which is an 
\emph{entangled} state of the constituent systems, or if it is a genuine 
\emph{mixed} state of the composite system, then the state of the 
constituent systems would necessarily be \emph{mixed}; moreover the 
empirical distance $d_{\bi\bj}$ would depend on the state of the subsystems 
other than the $\bi$'s and the $\bj$'s. Hence, in these cases the 
interpretation of $d_{\bi\bj}$ would not be obvious. Therefore, in the 
present paper, we assume that the states of the composite system are 
\emph{tensor products of pure vector states} of the constituent systems; 
and hence, without loss of generality, the composite system could be assumed 
to consist only of two subsystems. 

In this subsection, we calculate $d_{12}$ using (\ref{eq:4.2.1}) and discuss 
its properties at the genuine quantum level. The classical limit will be 
considered in subsection \ref{sub-4.4}. 

Let us write $\phi_1=\exp(-{\rm i}p_{1e}\xi^e_1/\hbar){\bf U}_1\psi_1$, where 
${\bf U}_1$ is the unitary operator on ${\cal H}_{P_1,s_1}$ representing an 
$SU(2)$ matrix $U^A_1{}_B$ and $\xi^e_1$ is a translation. Or, in other words, 
$\phi_1$ is considered to be obtained from the state $\psi_1$ by some $E(3)$ 
transformation. The state $\phi_2$ is assumed to have the analogous form. 
Then by (\ref{eq:4.1.2b}), (\ref{eq:3.1.6}) and (\ref{eq:3.1.7}) the first 
term in the numerator in (\ref{eq:4.2.1}) is 
\begin{eqnarray}
\langle\phi_1\otimes\phi_2\vert{\bf W}_{12}\vert\phi_1\otimes\phi_2\rangle
  \!\!\!\!&{}\!\!\!\!&=\bigl(\xi^a_1-\xi^a_2\bigr)\varepsilon_{abc}(R_1)^b{}_d
  \langle\psi_1\vert{\bf p}^d_1\vert\psi_1\rangle(R_2)^c{}_e\langle\psi_2\vert
  {\bf p}^e_2\vert\psi_2\rangle+ \nonumber \\
+\!\!\!\!&{}\!\!\!\!&(R^{-1}_1R_2)_{ab}\Bigl(\langle\psi_1\vert{\bf p}^a_1\vert
  \psi_1\rangle\langle\psi_2\vert{\bf J}^b_2\vert\psi_2\rangle+\langle\psi_1
  \vert{\bf J}^a_1\vert\psi_1\rangle\langle\psi_2\vert{\bf p}^b_2\vert\psi_2
  \rangle\Bigr). \label{eq:4.2.2}
\end{eqnarray}
In a similar way, the relevant factor in the second term of the numerator 
and the non-trivial term in the denominator, respectively, are 
\begin{eqnarray}
&{}&\langle\phi_1\otimes\phi_2\vert{\bf P}^2_{12}\vert\phi_1\otimes\phi_2
  \rangle=(R^{-1}_1R_2)_{ab}\langle\psi_1\vert{\bf p}^a_1\vert\psi_1\rangle
  \langle\psi_2\vert{\bf p}^b_2\vert\psi_2\rangle; \label{eq:4.2.3a} \\
&{}&\langle\phi_1\otimes\phi_2\vert{\bf P}^4_{12}\vert\phi_1\otimes\phi_2
  \rangle=(R^{-1}_1R_2)_{ab}(R^{-1}_1R_2)_{cd}\langle\psi_1\vert{\bf p}^a_1
  {\bf p}^c_1\vert\psi_1\rangle\langle\psi_2\vert{\bf p}^b_2{\bf p}^d_2\vert
  \psi_2\rangle. \label{eq:4.2.3b}
\end{eqnarray}
Thus, it is only $\xi^a_1-\xi^a_2$, i.e. only the \emph{relative} `position' 
of the two systems, that matters in the empirical distance. We will see that, 
in a similar way, it is only the \emph{relative} `orientation' of the two 
subsystems that matters. Next we specify the states $\psi_1$ and $\psi_2$. 

If $\psi={}_sY_{jm}/P$, then by (\ref{eq:A.2.5b}), (\ref{eq:A.2.6b}), 
(\ref{eq:A.2.7b}) and (\ref{eq:A.2.9a})-(\ref{eq:A.2.9c}) 
\begin{equation}
\langle\psi\vert{\bf p}^a\vert\psi\rangle=\delta^a_3P\frac{ms}{j(j+1)}, 
\hskip 20pt
\langle\psi\vert{\bf J}^a\vert\psi\rangle=\delta^a_3\hbar m. \label{eq:4.2.4}
\end{equation}
Thus, roughly speaking, in the state ${}_sY_{jm}/P$ both the linear and 
angular momenta point in the `$z$-direction' (with respect to the basis in 
the \emph{momentum space}). Using (\ref{eq:A.2.5a})-(\ref{eq:A.2.7c}), a 
direct calculation gives that 
\begin{eqnarray}
\langle{}_sY_{jm}\vert{\bf p}^1{\bf p}^1\vert{}_sY_{jm}\rangle\!\!\!\!&{}
  \!\!\!\!&=\langle{}_sY_{jm}\vert{\bf p}^2{\bf p}^2\vert{}_sY_{jm}\rangle=
  \frac{P^4}{j(j+1)(2j-1)(2j+3)}\Bigl(-3s^2m^2+ \nonumber \\
\!\!\!\!&{}\!\!\!\!&+j(j+1)\bigl(s^2+m^2\bigr)+j(j+1)\bigl(j^2+j-1\bigr)
  \Bigr), \label{eq:4.2.5a}\\
\langle{}_sY_{jm}\vert{\bf p}^3{\bf p}^3\vert{}_sY_{jm}\rangle\!\!\!\!&{}
  \!\!\!\!&=\frac{P^4}{j(j+1)(2j-1)(2j+3)}\Bigl(6s^2m^2-\nonumber \\
  \!\!\!\!&{}\!\!\!\!&-2j(j+1)\bigl(s^2+m^2\bigr)+j(j+1)\bigl(2j^2+2j-1\bigr)
  \Bigr); \label{eq:4.2.5b}
\end{eqnarray}
and that all the other components of $\langle{}_sY_{jm}\vert{\bf p}^a{\bf p}^b
\vert{}_sY_{jm}\rangle$ are vanishing. (\ref{eq:4.2.5a}) and (\ref{eq:4.2.5b}) 
imply that $\delta_{ab}\langle
{}_sY_{jm}\vert{\bf p}^a{\bf p}^b\vert{}_sY_{jm}\rangle=P^4$, as it should be. 
These expectation values may appear to be singular when $j=0$ or $1/2$, but 
these are not. In fact, in these cases $s=0$ and $\vert s\vert=1/2$, and 
hence $j=n$ and $j=1/2+n$, $n=0,1,2,...$, respectively. Writing these into 
(\ref{eq:4.2.5a}) and (\ref{eq:4.2.5b}) and then substituting $n=0$, we 
obtain that these are $P^4/3$ in both cases. 

Choosing both $\psi_1$ and $\psi_2$ in the above way, and substituting 
(\ref{eq:4.2.4}) into (\ref{eq:4.2.2}) and (\ref{eq:4.2.3a}), we find, 
respectively, that 
\begin{eqnarray*}
\langle\phi_1\otimes\phi_2\vert{\bf W}_{12}\vert\phi_1\otimes\phi_2\rangle=
  \!\!\!\!&{}\!\!\!\!&\bigl(\xi^a_1-\xi^a_2\bigr)\varepsilon_{abc}(R_1)^b{}_3
  (R_2)^c{}_3P_1P_2\frac{s_1s_2m_1m_2}{j_1(j_1+1)j_2(j_2+1)}+ \\
+\!\!\!\!&{}\!\!\!\!&(R^{-1}_1R_2)_{33}\hbar m_1m_2\Bigl(\frac{P_1s_1}{j_1
  (j_1+1)}+\frac{P_2s_2}{j_2(j_2+1)}\Bigr), \\
\langle\phi_1\otimes\phi_2\vert{\bf P}^2_{12}\vert\phi_1\otimes\phi_2\rangle
  =\!\!\!\!&{}\!\!\!\!&(R^{-1}_1R_2)_{33}P_1P_2\frac{s_1s_2m_1m_2}{j_1(j_1+1)
  j_2(j_2+1)}. 
\end{eqnarray*}
Hence, the numerator of (\ref{eq:4.2.1}) is 
\begin{eqnarray}
\langle\phi_1\otimes\phi_2\!\!\!\!&{}\!\!\!\!&\vert{\bf W}_{12}-\hbar\bigl(
  \frac{s_1}{P_1}+\frac{s_2}{P_2}\bigr){\bf P}^2_{12}\vert\phi_1\otimes\phi_2
  \rangle= \label{eq:4.2.6} \\
=\!\!\!\!&{}\!\!\!\!&P_1P_2\Bigl\{\bigl(\xi^a_1-\xi^a_2\bigr)\varepsilon_{abc}
  (R_1)^b{}_3(R_2)^c{}_3+  \nonumber \\
\!\!\!\!&{}\!\!\!\!&+(R^{-1}_1R_2)_{33}\hbar\Bigl(\frac{j_2(j_2+1)-s^2_2}
  {s_2P_2}+\frac{j_1(j_1+1)-s^2_1}{s_1P_1}\Bigr)\Bigr\}\frac{s_1s_2m_1m_2}
  {j_1(j_1+1)j_2(j_2+1)}. \nonumber
\end{eqnarray}
If $s_1s_2m_1m_2=0$, then this is zero. In this case at least one of the 
expectation values $\langle\phi_1\vert{\bf p}^a_1\vert\phi_1\rangle$ and 
$\langle\phi_2\vert{\bf p}^a_2\vert\phi_2\rangle$ is vanishing (see the 
first of (\ref{eq:4.2.4})). This case is analogous to the classical 
situation when $p^a_1=0$ or $p^a_2=0$, and that we excluded from our 
investigations (see the second paragraph in subsection \ref{sub-2.1}). It 
might be worth noting that (\ref{eq:4.2.6}) is just ($P_1P_2$ times) the 
expectation value of the distance \emph{operator} (\ref{eq:4.1.5}) in the 
zeroth approximation according to the expansion (\ref{eq:4.1.6}). 

Using $\langle{}_sY_{jm}\vert{\bf p}^1{\bf p}^1\vert{}_sY_{jm}\rangle=\langle
{}_sY_{jm}\vert{\bf p}^2{\bf p}^2\vert{}_sY_{jm}\rangle$, we obtain that, in 
the states above, (\ref{eq:4.2.3b}) takes the form 
\begin{eqnarray}
\langle\phi_1\otimes\phi_2\vert{\bf P}^4_{12}\vert\phi_1\otimes\phi_2\rangle=
  \!\!\!\!&{}\!\!\!\!&\Bigl(\bigl((R^{-1}_1R_2)_{11}\bigr)^2+\bigl((R^{-1}_1
  R_2)_{12}\bigr)^2+\bigl((R^{-1}_1R_2)_{21}\bigr)^2+ \nonumber\\
\!\!\!\!&{}\!\!\!\!&+\bigl((R^{-1}_1R_2)_{22}\bigr)^2\Bigr)\langle\psi_1\vert
  {\bf p}^1_1{\bf p}^1_1\vert\psi_1\rangle\langle\psi_2\vert{\bf p}^1_2
  {\bf p}^1_2\vert\psi_2\rangle+ \nonumber\\
+\!\!\!\!&{}\!\!\!\!&\Bigl(\bigl((R^{-1}_1R_2)_{13}\bigr)^2+\bigl((R^{-1}_1R_2)
  _{23}\bigr)^2\Bigr)\langle\psi_1\vert{\bf p}^1_1{\bf p}^1_1\vert\psi_1\rangle
  \langle\psi_2\vert{\bf p}^3_2{\bf p}^3_2\vert\psi_2\rangle+ \nonumber\\
+\!\!\!\!&{}\!\!\!\!&\Bigl(\bigl((R^{-1}_1R_2)_{31}\bigr)^2+\bigl((R^{-1}_1R_2)
  _{32}\bigr)^2\Bigr)\langle\psi_1\vert{\bf p}^3_1{\bf p}^3_1\vert\psi_1\rangle
  \langle\psi_2\vert{\bf p}^1_2{\bf p}^1_2\vert\psi_2\rangle+ \nonumber\\
+\!\!\!\!&{}\!\!\!\!&\bigl((R^{-1}_1R_2)_{33}\bigr)^2\langle\psi_1\vert{\bf p}
  ^3_1{\bf p}^3_1\vert\psi_1\rangle\langle\psi_2\vert{\bf p}^3_2{\bf p}^3_2
  \vert\psi_2\rangle. \label{eq:4.2.7}
\end{eqnarray}
If the $SU(2)$ matrix $U^A{}_B$ is parameterized by the familiar Euler angles 
$(\alpha,\beta,\gamma)$ according to 
\begin{equation}
U^A{}_B=\left(\begin{array}{ccc}
 \exp\bigl(\frac{\rm i}{2}(\alpha+\gamma)\bigr)\cos(\beta/2) &
 {\rm i}\exp\bigl(-\frac{\rm i}{2}(\alpha-\gamma)\bigr)\sin(\beta/2)\\
 {\rm i}\exp\bigl(\frac{\rm i}{2}(\alpha-\gamma)\bigr)\sin(\beta/2) &
 \exp\bigl(-\frac{\rm i}{2}(\alpha+\gamma)\bigr)\cos(\beta/2) \\
             \end{array}\right), \label{eq:4.2.8}
\end{equation}
then the corresponding rotation matrix is 
\begin{equation}
R^a{}_b=\left(\begin{array}{ccc}
 \cos\alpha\cos\gamma-\sin\alpha\cos\beta\sin\gamma &
 -\sin\alpha\cos\gamma-\cos\alpha\cos\beta\sin\gamma &
 \sin\beta\sin\gamma \\
 \cos\alpha\sin\gamma+\sin\alpha\cos\beta\cos\gamma &
 -\sin\alpha\sin\gamma+\cos\alpha\cos\beta\cos\gamma &
 -\sin\beta\cos\gamma \\
 \sin\alpha\sin\beta & \cos\alpha\sin\beta & \cos\beta 
\end{array}\right). \label{eq:4.2.9}
\end{equation}
This yields, in particular, that 
\begin{equation}
\cos\beta_{12}:=(R^{-1}_1R_2)_{33}=\cos\beta_1\cos\beta_2+\cos(\gamma_1-
\gamma_2)\sin\beta_1\sin\beta_2;  \label{eq:4.2.10}
\end{equation}
and that $\beta_{12}$ is just the angle between the unit vectors $(R_1)^a
{}_3$ and $(R_2)^a{}_3$, too: $\cos\beta_{12}=\delta_{ab}(R_1)^a{}_3(R_2)^b
{}_3$. Also, the combinations of the matrix elements $((R^{-1}_1R)_{ab})^2$ 
in (\ref{eq:4.2.7}) are all expressions of $\cos^2\beta_{12}$ alone: 
\begin{eqnarray}
\langle\phi_1\otimes\phi_2\vert{\bf P}^4_{12}\vert\phi_1\otimes\phi_2\rangle=
  \!\!\!\!&{}\!\!\!\!&\bigl(1+\cos^2\beta_{12}\bigr)\langle\psi_1\vert
  {\bf p}^1_1{\bf p}^1_1\vert\psi_1\rangle\langle\psi_2\vert{\bf p}^1_2
  {\bf p}^1_2\vert\psi_2\rangle+ \nonumber \\
+\!\!\!\!&{}\!\!\!\!&\bigl(1-\cos^2\beta_{12}\bigr)\langle\psi_1\vert
  {\bf p}^1_1{\bf p}^1_1\vert\psi_1\rangle\langle\psi_2\vert{\bf p}^3_2
  {\bf p}^3_2\vert\psi_2\rangle+ \nonumber \\
+\!\!\!\!&{}\!\!\!\!&\bigl(1-\cos^2\beta_{12}\bigr)\langle\psi_1\vert
  {\bf p}^3_1{\bf p}^3_1\vert\psi_1\rangle\langle\psi_2\vert{\bf p}^1_2
  {\bf p}^1_2\vert\psi_2\rangle+ \nonumber \\
+\!\!\!\!&{}\!\!\!\!&\cos^2\beta_{12}\,\langle\psi_1\vert{\bf p}^3_1{\bf p}^3_1
  \vert\psi_1\rangle\langle\psi_2\vert{\bf p}^3_2{\bf p}^3_2\vert\psi_2
  \rangle. \label{eq:4.2.11}
\end{eqnarray}
Since the length of the vector $\varepsilon^a{}_{bc}(R_1)^b{}_3(R_2)^c{}_3$ 
in (\ref{eq:4.2.6}) is $\sin\beta_{12}$, (\ref{eq:4.2.11}) shows that 
$d_{12}$ depends only on $\beta_{12}$, i.e. on the \emph{relative} 
`orientation' of the two constituent systems, rather than the individual 
Euler angles $(\alpha_1,\beta_1,\gamma_1)$ and $(\alpha_2,\beta_2,\gamma_2)$. 

Denoting the denominator in (\ref{eq:4.2.1}) by $P_1P_2D$ and substituting 
(\ref{eq:4.2.5a}) and (\ref{eq:4.2.5b}) into (\ref{eq:4.2.11}), a lengthy 
but straightforward calculation gives that 
\begin{eqnarray}
D^2\!\!\!\!&{}\!\!\!\!&=1-\frac{1}{(2j_1-1)(2j_1+3)(2j_2-1)(2j_2+3)}\Bigl(
  5j_1(j_1+1)j_2(j_2+1)+ \label{eq:4.2.12} \\
\!\!\!\!&{}\!\!\!\!&\hskip 160pt+ (s^2_1-4)j_2(j_2+1)+(s^2_2-4)j_1(j_1+1)-
  3s^2_1s^2_2+3\Bigr) \nonumber \\
-\!\!\!\!&{}\!\!\!\!&\frac{\bigl(j_1(j_1+1)-3s^2_1\bigr)\bigl(j_2(j_2+1)-3
  s^2_2\bigr)}{(2j_1-1)(2j_1+3)(2j_2-1)(2j_2+3)}\Bigl(\frac{m^2_1}{j_1(j_1+1)}
  +\frac{m^2_2}{j_2(j_2+1)}-\frac{3m^2_1m^2_2}{j_1(j_1+1)j_2(j_2+1)}\Bigr)
  \nonumber \\
-\!\!\!\!&{}\!\!\!\!&\frac{\bigl(j_1(j_1+1)-3s^2_1\bigr)\bigl(j_1(j_1+1)-3m
  ^2_1\bigr)\bigl(j_2(j_2+1)-3s^2_2\bigr)\bigl(j_2(j_2+1)-3m^2_2\bigr)}{j_1
  (j_1+1)(2j_1-1)(2j_1+3)j_2(j_2+1)(2j_2-1)(2j_2+3)}\cos^2\beta_{12}. \nonumber
\end{eqnarray}
If $j_1$ and $j_2$ take their \emph{smallest} value, viz. $j_1=\vert s_1\vert$ 
and $j_2=\vert s_2\vert$, i.e. when $\psi_1$ and $\psi_2$ are 
\emph{centre-of-mass states} (see subsection \ref{sub-3.2}), then this 
expression reduces to 
\begin{equation}
D^2=\frac{2}{3}+\frac{\bigl(3m^2_1-j_1(j_1+1)\bigr)\bigl(3m^2_2-j_2(j_2+1)
\bigr)}{3(j_1+1)(2j_1+3)(j_2+1)(2j_2+3)}(1-3\cos^2\beta_{12}).
\label{eq:4.2.13}
\end{equation}
If at least one of $j_1=\vert s_1\vert$ and $j_2=\vert s_2\vert$ is $0$ or 
$1/2$, then $D^2=2/3$, and hence, in particular, it is not zero and it does 
not depend on $\beta_{12}$. For small spins the dependence of $D^2$ on 
$\beta_{12}$ is weak. The higher the spins $s_1$ and $s_2$, the closer the 
$D^2$ to zero for $\vert m_1\vert=j_1$, $\vert m_2\vert=j_2$ and $\cos^2
\beta_{12}=1$. Nevertheless, we show that $D^2$ is \emph{strictly positive} 
for any finite $\vert m_1\vert\leq j_1=\vert s_1\vert$ and $\vert m_2\vert
\leq j_2=\vert s_2\vert$ and any angle $\beta_{12}$. 

Suppose, on the contrary, that $D^2=0$, i.e. that for some $m_1$, $m_2$ and 
$\beta_{12}$ 
\begin{equation}
2(j_1+1)(j_2+1)(2j_1+3)(2j_2+3)=\bigl(3m^2_1-j_1(j_1+1)\bigr)\bigl(3m^2_2-
j_2(j_2+1)\bigr)(3\cos^2\beta_{12}-1) \label{eq:4.2.14}
\end{equation}
holds. Since the left hand side is positive, $3\cos^2\beta_{12}\not=1$, $3m_1
^2\not=j_1(j_1+1)$, $3m_2^2\not=j_2(j_2+1)$ and $j_1\not=0\not=j_2$ must hold. 
First, let us suppose that $-1\leq 3\cos^2\beta_{12}-1<0$. Then by 
(\ref{eq:4.2.14}) $3m^2_1>j_1(j_1+1)$ and $3m^2_2<j_2(j_2+1)$ or $3m^2_1<j_1
(j_1+1)$ and $3m^2_2>j_2(j_2+1)$ follow. Let us consider the first case. Then, 
also by (\ref{eq:4.2.14}), 
\begin{eqnarray*}
&{}&2(j_1+1)(j_2+1)\bigl(4j_1j_2+6(j_1+j_2)+9\bigr)\leq-\bigl(3m^2_1-j_1
  (j_1+1)\bigr)\bigl(3m^2_2-j_2(j_2+1)\bigr) \\
&{}&\hskip 10pt <3m_1^2j_2(j_2+1)-3m^2_2\bigl(3m^2_1-j_1(j_1+1)\bigr)-j_1j_2
  (j_1+1)(j_2+1)<3m^2_1j_2(j_2+1) \\
&{}&\hskip 10pt \leq3j^2_1j_2(j_2+1).
\end{eqnarray*}
This implies $8j^2_1j_2<2(j_1+1)(4j_1j_2+6j_1+6j_2+9)\leq3j^2_1j_2$, which is 
a contradiction. The proof is similar if $3m^2_1<j_1(j_1+1)$ and $3m^2_2>j_2
(j_2+1)$. Next suppose that $0<3\cos^2\beta_{12}-1\leq2$. Then 
(\ref{eq:4.2.14}) implies that
\begin{equation*}
(j_1+1)(j_2+1)\bigl(4j_1j_2+6j_1+6j_2+9\bigr)\leq\bigl(3m^2_1-j_1(j_1+1)\bigr)
\bigl(3m^2_2-j_2(j_2+1)\bigr),
\end{equation*}
and that either $3m^2_1>j_1(j_1+1)$ and $3m^2_2>j_2(j_2+1)$ or $3m^2_1<j_1
(j_1+1)$ and $3m^2_2<j_2(j_2+1)$. In the first case this yields 
\begin{eqnarray*}
&{}&(j_1+1)(j_2+1)\bigl(3j_1j_2+6j_1+6j_2+9\bigr)\leq9m^2_1m^2_2-3m^2_1j_2
  (j_2+1)-3m^2_2j_1(j_1+1) \\
&{}&\hskip 20pt \leq9m^2_1m^2_2-3m^2_1j^2_2-3m^2_2j^2_1\leq3m^2_1m^2_2-3m^2_1
  (j^2_2-m^2_2)-3m^2_2(j^2_1-m^2_1) \\
&{}&\hskip 20pt \leq3m^2_1m^2_2\leq 3j^2_1j^2_2,
\end{eqnarray*}
which is a contradiction. In the second case, 
\begin{eqnarray*}
&{}&(j_1+1)(j_2+1)\bigl(3j_1j_2+6j_1+6j_2+9\bigr)\leq9m^2_1m^2_2-3m^2_1j_2
  (j_2+1)-3m^2_2j_1(j_1+1) \\
&{}&\hskip 20pt <9m^2_1m^2_2-9m^2_1m^2_2-9m^2_1m^2_2=-9m^2_1m^2_2,
\end{eqnarray*}
which is also a contradiction. We expect that the denominator $D$, given by 
(\ref{eq:4.2.12}), is not zero even in the general case when $j_1>\vert s_1
\vert$ and $j_2>\vert s_2\vert$. 

The other extreme case is when both $j_1$ and $j_2$ tend to infinity. Now 
there are three sub-cases: when $m_1$ and $m_2$ remain bounded, and when 
one of them, say $m_1$, or both tend to infinity with $j_1$ and $j_2$. As 
equations (\ref{eq:4.2.4}) show, in all these cases the expectation value 
of the linear momenta tends to zero, but in the first case the expectation 
value of the angular momenta remain finite; in the second the expectation 
value of ${\bf J}^a_1$ diverges but that of ${\bf J}^a_2$ remains finite; 
while in the third the expectation value of ${\bf J}^a_1$ and ${\bf J}^a_2$ 
diverges. By (\ref{eq:4.2.12}) these limits of $D^2$ are 
\begin{equation}
\frac{11}{16}-\frac{1}{16}\cos^2\beta_{12}, \hskip 25pt 
\frac{5}{8}+\frac{1}{8}\cos^2\beta_{12}, \hskip 25pt
\frac{3}{4}-\frac{1}{4}\cos^2\beta_{12}, \label{eq:4.2.15}
\end{equation}
respectively. These are independent of the spins, and none of them is zero. 

Therefore, the empirical distance $d_{12}$ between the elementary systems is 
well defined, finite or zero, at least in the states obtained from 
centre-of-mass states by $E(3)$ transformations, and also in the $j_1,j_2\to
\infty$ limit; and depends only on the \emph{relative} `position' and 
`orientation' of the subsystems. Next we discuss the resulting expression of 
$d_{12}$ in these two extreme cases. 

In the first case, i.e. when $j_1=\vert s_1\vert$ and $j_2=\vert s_2\vert$,  
\begin{equation}
d_{12}=\frac{1}{D}\Bigl((\xi^a_1-\xi^a_2)\varepsilon_{abc}(R_1)^b{}_3(R_2)^c{}_3
+\hbar\bigl(\frac{s_1}{\vert s_1\vert P_1}+\frac{s_2}{\vert s_2\vert P_2}\bigr)
\cos\beta_{12}\Bigr)\frac{s_1s_2}{\vert s_1s_2\vert}\frac{m_1m_2}{(\vert s_1
\vert+1)(\vert s_2\vert+1)}, \label{eq:4.2.16}
\end{equation}
where now $D$ is given by (\ref{eq:4.2.13}), and the components of the unit 
vectors $(R_1)^a{}_3$ and $(R_2)^a{}_3$ can be read off from (\ref{eq:4.2.9}). 
The first term in the brackets can be zero when these unit vectors are 
parallel, $(R_1)^a{}_3=\pm(R_2)^a{}_3$, i.e. when $\beta_{12}=0$ or $\pi$, or 
when $\xi^a_1-\xi^a_2$ is zero or at least it lays in the 2-plane spanned by 
$(R_1)^a{}_3$ and $(R_2)^a{}_3$. This term can be arbitrarily large, depending 
on $\xi^a_1-\xi^a_2$. This does not contain Planck's constant and the Casimir 
invariants $P_1$ and $P_2$, and gives the `classical part' of the distance, 
being analogous to the last term on the right of equation (\ref{eq:2.2.1}). 

The second term in the brackets, being proportional to $\hbar$, is a genuine 
quantum correction to the classical part. $d_{12}$ depends on $P_1$ and $P_2$ 
only through this term. This can be zero only if $\beta_{12}=\pi/2$; and, for 
$\beta_{12}\not=\pi/2$, only in the \emph{very exceptional case} when $P_1=
P_2$ and ${\rm sign}(s_1)=-{\rm sign}(s_2)$, i.e. if one of $\psi_1$ and 
$\psi_2$ is holomorphic and the other is anti-holomorphic. 

Even if $\xi^a_1-\xi^a_2$, $(R_1)^a{}_3$ and $(R_2)^a{}_3$ are given, $d_{12}$ 
is not fixed: it depends on the discrete `quantum numbers' $m_1$ and $m_2$ 
of the actual states in an essential way. In particular, for $s_1=s_2=1/2$ 
(\ref{eq:4.2.16}) gives 
\begin{equation*}
d_{12}=\pm\frac{1}{3\sqrt{6}}\Bigl(\bigl(\xi^a_1-\xi^a_2\bigr)\varepsilon
_{abc}(R_1)^a{}_3(R_2)^c{}_3+\hbar\frac{P_1+P_2}{P_1P_2}\cos\beta_{12}\Bigr).
\end{equation*}
Its `classical part' is less than one-sixth of the distance between the two 
classical point particles characterized by the same classical Euclidean 
transformations $((R_1)^a{}_b,\xi^a_1)$ and $((R_2)^a{}_b,\xi^a_2)$. 

If $\cos\beta_{12}=\pm1$, then the first term between the brackets in 
(\ref{eq:4.2.16}) vanishes, and $d_{12}$ becomes an expression of the 
Casimir invariants $(P_1,s_1)$, $(P_2,s_2)$ of the elementary systems and 
the discrete quantum numbers $m_1$ and $m_2$ alone. So the distance in this 
case is `universal', it is of purely quantum mechanical origin, and, apart 
from the very exceptional case above, non-zero. Thus, at the quantum level, 
the expression for $d_{12}$ is well defined, in contrast to the classical 
case when, by the discussion of subsection \ref{sub-2.2}, the distance 
between two centre-of-mass lines with parallel linear momenta could be 
recovered only as a limit. 

As we concluded above, $d_{12}$ is well defined also in the other extreme 
case when $s_1$ and $s_2$ are fixed but $j_1,j_2\to\infty$. Now we determine 
the distance in this case explicitly. As (\ref{eq:4.2.6}) shows, in the 
first two cases considered in (\ref{eq:4.2.15}) the empirical distance 
$d_{12}$ tends to zero, while in the third (i.e. when $\vert m_1\vert=j_1$, 
$\vert m_2\vert=j_2\to\infty$) it tends to
\begin{equation*}
2\hbar\bigl(\frac{s_1}{P_2}+\frac{s_2}{P_1}\bigr)\frac{\cos\beta_{12}}{\sqrt{
3-\cos^2\beta_{12}}}.
\end{equation*}
In particular, this limit is independent of $\xi^a_1-\xi^a_2$, and $d_{12}$ 
reduces to the quantum correction. Nevertheless, this extreme case 
corresponds to a rather exotic situation, since by (\ref{eq:4.2.4}) the 
expectation value of the angular momenta tend to infinity while that of the 
linear momenta to zero. 

As we noted, in the states with the choice for $\psi_1$, $\psi_2$ above all 
the expectation values are vanishing for $s_1=0$, $s_2=0$. To get non-zero 
results in this case, more general states with $\psi=\sum_{j,m}c^{jm}{}_s
Y_{jm}$ should be considered, because the matrix elements $\langle{}_s
Y_{j\pm1,n}\vert{\bf p}^a\vert{}_sY_{j,m}\rangle$ are not all zero even for 
$s=0$ (see Appendix \ref{sub-A.2}). However, the explicit form of the 
resulting expectation values are much more complicated. 


\subsection{The empirical angles and volume elements}
\label{sub-4.3}

Dictated by the classical formula (\ref{eq:2.3.1}), we define the `empirical 
angle' between the linear momenta of two \emph{elementary} subsystems 
(characterized by $(P_1,s_1)$ and $(P_2,s_2)$, respectively) in their pure 
tensor product state $\phi=\phi_1\otimes\phi_2$ by 
\begin{equation}
\cos\omega_{12}:=\frac{\langle\phi\vert\delta_{ab}{\bf p}^a_1\otimes{\bf p}^b_2
\vert\phi\rangle}{\sqrt{\langle\phi\vert{\bf P}^2_1\vert\phi\rangle}\sqrt{
\langle\phi\vert{\bf P}^2_2\vert\phi\rangle}}=\frac{\langle\phi_1\vert{\bf p}
^a_1\vert\phi_1\rangle\delta_{ab}\langle\phi_2\vert{\bf p}^b_2\vert\phi_2
\rangle}{P_1P_2} \label{eq:4.3.1}
\end{equation}
with range $\omega_{12}\in[0,\pi]$. 
Also, motivated by the classical expression (\ref{eq:2.3.2}), we define the 
`empirical 3-volume element' for three elementary systems in the pure tensor 
product state $\phi=\phi_1\otimes\phi_2\otimes\phi_3$ by 
\begin{equation}
v_{123}:=\frac{1}{3!}\varepsilon_{abc}\frac{\langle\phi\vert{\bf p}^a_1\otimes
{\bf p}^b_2\otimes{\bf p}^c_3\vert\phi\rangle}{\sqrt{\langle\phi\vert{\bf P}
^2_1\vert\phi\rangle}\sqrt{\langle\phi\vert{\bf P}^2_2\vert\phi\rangle}
\sqrt{\langle\phi\vert{\bf P}^2_3\vert\phi\rangle}}. \label{eq:4.3.2}
\end{equation}
Since these quantities are built only from the momentum operators, and the 
momentum operators are invariant with respect to translations, it is enough 
to evaluate these expressions only in the states of the form $\phi={\bf U}
\psi$. 

Thus, if $\phi_1={\bf U}_1\psi_1$ and $\phi_2={\bf U}_2\psi_2$, then 
\begin{eqnarray}
\cos\omega_{12}\!\!\!\!&=\!\!\!\!&\frac{1}{P_1P_2}\langle\psi_1\vert{\bf U}
  ^\dagger_1{\bf p}^a_1{\bf U}_1\vert\psi_1\rangle\delta_{ab}\langle\psi_2\vert
  {\bf U}^\dagger_2{\bf p}^b_2{\bf U}_2\vert\psi_2\rangle= \nonumber \\
\!\!\!\!&=\!\!\!\!&\frac{1}{P_1P_2}(R^{-1}_1R_2)_{ab}\langle\psi_1\vert{\bf p}
  ^a_1\vert\psi_1\rangle\langle\psi_2\vert{\bf p}^b_2\vert\psi_2\rangle.
\label{eq:4.3.3}
\end{eqnarray}
In particular, if $\psi_1={}_{s_1}Y_{j_1,m_1}/P_1$ and $\psi_2={}_{s_2}Y_{j_2,m_2}/
P_2$, then 
\begin{equation}
\cos\omega_{12}=s_1s_2\frac{m_1m_2}{j_1(j_1+1)j_2(j_2+1)}\cos\beta_{12}.
\label{eq:4.3.4}
\end{equation}
The angle $\omega_{12}$ has the same qualitative properties that the 
empirical angle $\theta_{12}$ has in the $SU(2)$-invariant systems 
\cite{Sz21c}. In particular, for given $s_1$ and $s_2$ and angle $\beta_{12}$, 
the empirical angle $\omega_{12}$ is still not fixed, that may take different 
\emph{discrete} values. Moreover, $\omega_{12}$ is \emph{never} zero even if 
$\beta_{12}=0$, and is never $\pi$ even if $\beta_{12}=\pi$. With given $s_1$ 
and $s_2$ the empirical angle $\omega_{12}$ takes its minimal value in the 
special centre-of-mass states when $m_1=j_1=\vert s_1\vert$ and $m_2=j_2=
\vert s_2\vert$, and $\beta_{12}=0$. In particular, for $\vert s_1\vert=
\vert s_2\vert=1/2$ this angle is $\omega^{min}_{12}\approx83.62^\circ$, while 
for $\vert s_1\vert=\vert s_2\vert=1$ it is $\approx75.52^\circ$. The maximal 
value of $\omega_{12}$ is $\omega^{max}_{12}=\pi-\omega^{min}_{12}$. This 
minimal/maximal value tends to zero/$\pi$ only in the $\vert s_1\vert$, 
$\vert s_2\vert\to\infty$ limit. (For a classical model of the `geometry of 
the quantum directions', see Section 4 of \cite{Sz21c}.) 

The evaluation of the empirical 3-volume element in the analogous states is 
similar. By the first of (\ref{eq:4.2.4}) we obtain 
\begin{equation}
v_{123}=\frac{s_1s_2s_3m_1m_2m_3}{j_1(j_1+1)j_2(j_2+1)j_3(j_3+1)}\frac{1}{3!}
\varepsilon_{abc}(R_1)^a{}_3(R_2)^b{}_3(R_3)^c{}_3. \label{eq:4.3.5}
\end{equation}
The second factor in (\ref{eq:4.3.5}) is just the Euclidean 3-volume of the 
tetrahedron spanned by the unit vectors $(R_1)^a{}_3$, $(R_2)^a{}_3$ and 
$(R_3)^a{}_3$. Hence, even for given $(R_1)^a{}_3$, $(R_2)^a{}_3$ and $(R_3)^a
{}_3$, the empirical 3-volume element takes different \emph{discrete} values, 
depending on the `quantum numbers' of the states of the constituent 
subsystems. By (\ref{eq:4.3.5}) $v_{123}$ is \emph{always} smaller than the 
Euclidean 3-volume element, even if all the $j$'s take their minimal, and 
all the $m$'s take their maximal value, viz. $m=j=\vert s\vert$. 


\subsection{The classical limit of the empirical geometrical quantities}
\label{sub-4.4}

Traditionally, the classical limit of an $SU(2)$-invariant 
system is defined to be the limit in which $m=j\to\infty$ (see e.g. 
\cite{Wi}). However, by $\vert m\vert\leq j$ and (\ref{eq:4.2.4}), in the 
present $E(3)$-invariant case, the expectation value of the linear momentum 
tends to zero unless the spin $s$ also tends to infinity; and this 
expectation value can tend to a large macroscopic value if $P$ is also 
growing appropriately. Thus, formally, the classical limit of the 
$E(3)$-invariant systems should be defined to be the limit in which 
$\vert s\vert=m=j\to\infty$ and $P\to\infty$. Moreover, if we expect that 
the expectation value of the linear and angular momenta tend to the 
corresponding large classical value in the same order, then by 
(\ref{eq:4.2.4}) we should require that asymptotically $P=p j+O(1/j)$ 
holds for some positive $p$. We use this latter condition in the calculation 
of the classical limit of the uncertainty of the empirical distance. Note 
that by $j=\vert s\vert$ the states in such a sequence are obtained from 
\emph{centre-of-mass states} by some $E(3)$ transformations. In the present 
subsection, we determine this limit of the empirical distances and their 
uncertainty, and also that of the angles and 3-volume elements. We find that 
these are just those in the Euclidean 3-space. 

The $s_1=m_1=j_1$, $\pm s_2=m_2=j_2\to\infty$ limit of the numerator in the 
expression (\ref{eq:4.2.16}) of the empirical distance is 
\begin{equation*}
\pm(\xi^a_1-\xi^a_2)\varepsilon_{abc}(R_1)^b{}_3(R_2)^c{}_3\pm\hbar
\frac{P_1\pm P_2}{P_1P_2}\cos\beta_{12};
\end{equation*}
while, by (\ref{eq:4.2.13}), the same limit of its denominator is $\sin
\beta_{12}$. But the latter is just the magnitude of $\varepsilon^a{}_{bc}
(R_1)^b{}_3(R_2)^c{}_3$, and hence, apart from the quantum correction, for 
$\beta_{12}\not=0,\pi$ this gives just the classical (signed) distance 
between the centre-of-mass lines of the two classical point particles 
characterized by $((R_1)^a{}_b,\xi^a_1)$ and $((R_2)^a{}_b,\xi^a_2)$, 
respectively. If, in addition, $P_1,P_2\to\infty$, then the quantum 
correction goes away, and the whole expression reduces to the classical 
empirical distance. 

Since $d_{12}$ is \emph{not} the expectation value of some quantum observable, 
its standard deviation/variance cannot be defined in the standard manner. 
Nevertheless, its uncertainty can be introduced by importing the idea 
from experimental physics how the error of a quantity, built from 
experimental data, is defined. Namely, if $Q=Q(q^1,...,q^n)$ is a 
differentiable function of its variables and, in a series of measurements, 
we obtain the mean values $\bar q^\alpha$ and errors $\delta q^\alpha$ of the 
quantities $q^\alpha$, $\alpha=1,...,n$, then the mean value and error of 
$Q$ are defined to be $\bar Q:=Q(\bar q^1,...,\bar q^n)$ and 
\begin{equation*}
\delta Q:=\vert\frac{\partial Q}{\partial q^1}(\bar q^\alpha)\vert\delta q^1+
\cdots+\vert\frac{\partial Q}{\partial q^n}(\bar q^\alpha)\vert\delta q^n,
\end{equation*}
respectively. Hence, since the empirical distance has the structure 
\begin{equation*}
d_{12}=\frac{\langle\phi\vert{\bf A}_{12}\vert\phi\rangle}{\sqrt{\langle\phi
\vert{\bf B}_{12}\vert\phi\rangle}}
\end{equation*}
with ${\bf A}_{12}:={\bf W}_{12}-\hbar(s_1/P_1+s_2/P_2){\bf P}^2_{12}$ and 
${\bf B}_{12}:=P^2_1P^2_2-{\bf P}^4_{12}$, it seems natural to define the 
\emph{uncertainty} of $d_{12}$ in the state $\phi=\phi_1\otimes\phi_2$ to be 
\begin{equation}
\delta_\phi d_{12}:=\Bigl(\frac{\Delta_\phi{\bf A}_{12}}{\vert\langle\phi\vert
{\bf A}_{12}\vert\phi\rangle\vert}+\frac{1}{2}\frac{\Delta_\phi{\bf B}_{12}}
{\langle\phi\vert{\bf B}_{12}\vert\phi\rangle}\Bigr)\vert d_{12}\vert. 
\label{eq:4.3.6}
\end{equation}
Here $(\Delta_\phi{\bf A}_{12})^2=\langle\phi\vert({\bf A}_{12})^2\vert\phi
\rangle-(\langle\phi\vert{\bf A}_{12}\vert\phi\rangle)^2$ and $(\Delta_\phi
{\bf B}_{12})^2=\langle\phi\vert({\bf B}_{12})^2\vert\phi\rangle-(\langle\phi
\vert{\bf B}_{12}\vert\phi\rangle)^2$, the two familiar variances. We are 
going to show that, in the classical limit defined above, both terms between 
the brackets tend to zero. Since $\vert d_{12}\vert$ in this limit is bounded, 
this means that the uncertainty $\delta_\phi d_{12}$ also tends to zero. 

Since the subsequent calculations are quite lengthy but elementary, we do 
not provide all the details. We indicate only the key steps. First, let us 
consider $\langle\phi\vert({\bf A}_{12})^2\vert\phi\rangle=\langle\phi\vert(
{\bf W}^2_{12}-2\hbar(j_1/P_1\pm j_2/P_2){\bf P}^2_{12}{\bf W}_{12}+\hbar^2(j_1
/P_1\pm j_2/P_2)^2{\bf P}^4_{12})\vert\phi\rangle$. The expectation values in 
\begin{eqnarray*}
\langle\phi_1\otimes\phi_2\vert{\bf W}^2_{12}\vert\phi_1\otimes\phi_2\rangle
  \!\!\!\!&=\!\!\!\!&\langle\phi_1\vert{\bf p}^a_1{\bf p}^b_1\vert\phi_1
  \rangle\delta_{ac}\delta_{bd}\langle\phi_2\vert{\bf J}^c_2{\bf J}^d_2\vert
  \phi_2\rangle+\langle\phi_1\vert{\bf p}^a_1{\bf J}^b_1\vert\phi_1
  \rangle\delta_{ac}\delta_{bd}\langle\phi_2\vert{\bf J}^c_2{\bf p}^d_2\vert
  \phi_2\rangle \\
\!\!\!\!&+\!\!\!\!&\langle\phi_1\vert{\bf J}^a_1{\bf p}^b_1\vert\phi_1
  \rangle\delta_{ac}\delta_{bd}\langle\phi_2\vert{\bf p}^c_2{\bf J}^d_2\vert
  \phi_2\rangle+\langle\phi_1\vert{\bf J}^a_1{\bf J}^b_1\vert\phi_1
  \rangle\delta_{ac}\delta_{bd}\langle\phi_2\vert{\bf p}^c_2{\bf p}^d_2\vert
  \phi_2\rangle
\end{eqnarray*}
can be calculated by using (\ref{eq:3.1.6a}). These are 
\begin{eqnarray*}
\langle\phi\vert{\bf p}^a{\bf p}^b\vert\phi\rangle\!\!\!\!&=\!\!\!\!&R^a{}_c
  R^b{}_d\langle\psi\vert{\bf p}^c{\bf p}^d\vert\psi\rangle, \\
\langle\phi\vert{\bf p}^a{\bf J}^b\vert\phi\rangle\!\!\!\!&=\!\!\!\!&R^a{}_c
  R^b{}_d\langle\psi\vert{\bf p}^c{\bf J}^d\vert\psi\rangle+R^a{}_c\varepsilon
  ^b{}_{ef}\xi^eR^f{}_d\langle\psi\vert{\bf p}^c{\bf p}^d\vert\psi\rangle, \\
\langle\phi\vert{\bf J}^a{\bf J}^b\vert\phi\rangle\!\!\!\!&=\!\!\!\!&R^a{}_c
  R^b{}_d\langle\psi\vert{\bf J}^c{\bf J}^d\vert\psi\rangle+R^a{}_c\varepsilon
  ^b{}_{ef}\xi^eR^f{}_d\overline{\langle\psi\vert{\bf p}^d{\bf J}^c\vert\psi
  \rangle}+ \\
\!\!\!\!&+\!\!\!\!&\varepsilon^a{}_{ef}\xi^eR^f{}_cR^b{}_d\langle\psi{\bf p}^c
  {\bf J}^d\vert\psi\rangle+\varepsilon^a{}_{ef}\xi^eR^f{}_c\varepsilon^b{}_{gh}
  \xi^gR^h{}_d\langle\psi\vert{\bf p}^c{\bf p}^d\vert\psi\rangle,
\end{eqnarray*}
where some of the expectation values with $\psi={}_{\pm j}Y_{jj}/P$ have already 
been given by (\ref{eq:4.2.5a}) and (\ref{eq:4.2.5b}), while the others can 
be calculated by (\ref{eq:A.2.5a})-(\ref{eq:A.2.7c}) and 
(\ref{eq:A.2.9a})-(\ref{eq:A.2.9c}). These are 
\begin{eqnarray*}
\langle\psi\vert{\bf p}^a{\bf J}^b\vert\psi\rangle\!\!\!\!&=\!\!\!\!&\pm
  \frac{\hbar Pj}{2(j+1)}\Bigl(\delta^a_1\delta^b_1+\delta^a_2\delta^b_2+2j
  \delta^a_3\delta^b_3+{\rm i}\bigl(\delta^a_1\delta^a_2-\delta^a_2\delta^a_1
  \bigr)\Bigr), \\
\langle\psi\vert{\bf J}^a{\bf J}^b\vert\psi\rangle\!\!\!\!&=\!\!\!\!&
  \frac{1}{2}\hbar^2j\Bigl(\delta^a_1\delta^b_1+\delta^a_2\delta^b_2+2j
  \delta^a_3\delta^b_3+{\rm i}\bigl(\delta^a_1\delta^a_2-\delta^a_2\delta^a_1
  \bigr)\Bigr).
\end{eqnarray*}
Using (\ref{eq:4.2.10}) and that, for large $j$, asymptotically $P=pj+O
(1/j)$, we obtain 
\begin{eqnarray}
\frac{1}{P^2_1P^2_2}\langle\phi_1\otimes\phi_2\vert{\bf W}^2_{12}\vert\phi_1
  \otimes\phi_2\rangle\!\!\!\!&=\!\!\!\!&\Bigl(\bigl(\xi^a_1-\xi^a_2\bigr)
  \varepsilon_{abc}(R_1)^b{}_3(R_2)^c{}_3\Bigr)^2+\nonumber \\
\!\!\!\!&+\!\!\!\!&2\hbar\bigl(\frac{j_1}{P_1}\pm\frac{j_2}{p_2}\bigr)\bigl(
  \xi^a_1-\xi^a_2\bigr)\varepsilon_{abc}(R_1)^b{}_3(R_2)^c{}_3\cos\beta_{12}+
  \nonumber \\
\!\!\!\!&+\!\!\!\!&\hbar^2\bigl(\frac{j_1}{P_1}\pm\frac{j_2}{P_2}\bigr)^2
 \cos^2\beta_{12}+O(\frac{1}{j_1})+O(\frac{1}{j_2}). \label{eq:4.3.7a}
\end{eqnarray}
The calculation of the expectation value of the other terms is similar, and 
for them we obtain 
\begin{eqnarray}
\!\!\!\!&{}\!\!\!\!&\frac{1}{P^2_1P^2_2}\langle\phi_1\otimes\phi_2\vert-2\hbar
  \bigl(\frac{j_1}{P_1}\pm\frac{j_2}{P_2}\bigr){\bf P}^2_{12}{\bf W}^2_{12}+
  \hbar^2\bigl(\frac{j_1}{P_1}\pm\frac{j_2}{P_2}\bigr)^2{\bf P}^4_{12}\vert
  \phi_1\otimes\phi_2\rangle \label{eq:4.3.7b} \\
=\!\!\!\!&{}\!\!\!\!&-2\hbar\bigl(\frac{j_1}{P_1}\pm\frac{j_2}{P_2}\bigr)
  \bigl(\xi^a_1-\xi^a_2\bigr)\varepsilon_{abc}(R_1)^b{}_3(R_2)^c{}_3\cos
  \beta_{12}-\hbar^2\bigl(\frac{j_1}{P_1}\pm\frac{j_2}{P_2}\bigr)^2\cos^2
  \beta_{12}+O(\frac{1}{j_1})+O(\frac{1}{j_2}). \nonumber
\end{eqnarray}
Comparing this with (\ref{eq:4.3.7a}) and recalling that the first term on 
the right of (\ref{eq:4.3.7a}) is just the classical limit of the square of 
$\langle\phi_1\otimes\phi_2\vert{\bf W}_{12}-\hbar(j_1/P_1\pm j_2/P_2){\bf P}
^2_{12}\vert\phi_1\otimes\phi_2\rangle/P_1P_2$, we find that the first term in 
the brackets in (\ref{eq:4.3.6}) is of order $O(1/j_1)+O(1/j_2)$. 

Using (\ref{eq:4.2.11}), (\ref{eq:4.2.5a}) and (\ref{eq:4.2.5b}), we 
immediately obtain the asymptotic form of the expectation value of the first 
two terms in ${\bf B}^2_{12}/P^4_1P^4_2=1-2{\bf P}^4_{12}/P^2_1P^2_2+{\bf P}^8
_{12}/P^4_1P^4_2$. It is 
\begin{equation*}
\frac{1}{P^4_1P^4_2}\langle\phi_1\otimes\phi_2\vert P^4_1P^4_2-2P^2_1P^2_2
{\bf P}^4_{12}\vert\phi_1\otimes\phi_2\rangle=1-2\cos^2\beta_{12}+O(\frac{1}
{j_1})+O(\frac{1}{j_2}).
\end{equation*}
Thus, we should calculate only the asymptotic form of 
\begin{eqnarray*}
&{}&\langle\phi_1\otimes\phi_2\vert{\bf P}^8_{12}\vert\phi_1\otimes\phi_2
  \rangle=\langle\phi_1\vert{\bf p}^a_1{\bf p}^b_1{\bf p}^c_1{\bf p}^d_1\vert
  \phi_1\rangle\delta_{ae}\delta_{bf}\delta_{cg}\delta_{dh}\langle\phi_2
  \vert{\bf p}^e_2{\bf p}^f_2{\bf p}^g_2{\bf p}^h_2\vert\phi_2\rangle= \\
&{}&=\langle\psi_1\vert{\bf p}^a_1{\bf p}^b_1{\bf p}^c_1{\bf p}^d_1\vert\psi_1
  \rangle(R^{-1}_1R_2)_{ae}(R^{-1}_1R_2)_{bf}(R^{-1}_1R_2)_{cg}(R^{-1}_1R_2)_{dh}
  \langle\psi_2\vert{\bf p}^e_2{\bf p}^f_2{\bf p}^g_2{\bf p}^h_2\vert\psi_2
  \rangle
\end{eqnarray*}
with $\psi_1={}_{j_1}Y_{j_1,j_1}/P_1$ and $\psi_2={}_{\pm j_2}Y_{j_2,j_2}/P_2$ when 
$j_1,j_2\to\infty$. Since by (\ref{eq:A.2.5a})-(\ref{eq:A.2.7c}) 
\begin{equation*}
\langle{}_{\pm j}Y_{jj}\vert{\bf p}^3\vert{}_{\pm j}Y_{jj}\rangle=\pm P^3\frac{j}
{j+1}+O(\frac{1}{j})
\end{equation*}
and all the other matrix elements of the form $\langle{}_{\pm j}Y_{jj}\vert
{\bf p}^a\vert{}_{\pm j}Y_{lm}\rangle$ fall off as $1/\sqrt{j}$ for large $j$, 
we have that 
\begin{equation*}
\langle{}_{\pm j}Y_{jj}\vert{\bf p}^a{\bf p}^b{\bf p}^c{\bf p}^d\vert{}_{\pm j}
Y_{jj}\rangle=P^6\frac{j^4}{(j+1)^4}\delta^a_3\delta^b_3\delta^c_3\delta^d_3
+O(\frac{1}{\sqrt{j}}). 
\end{equation*}
Thus, by (\ref{eq:4.2.10}) 
\begin{equation*}
\frac{1}{P^4_1P^4_2}\langle\phi_1\otimes\phi_2\vert{\bf P}^8_{12}\vert\phi_1
\otimes\phi_2\rangle=\cos^4\beta_{12}+O(\frac{1}{\sqrt{j_1}})+O(\frac{1}
{\sqrt{j_2}}).
\end{equation*}
Hence 
\begin{eqnarray*}
\frac{1}{P^4_1P^4_2}\langle\phi_1\otimes\phi_2\vert{\bf B}^2_{12}\vert\phi_1
  \otimes\phi_2\rangle\!\!\!\!&=\!\!\!\!&\sin^4\beta_{12}+O(\frac{1}
  {\sqrt{j_1}})+O(\frac{1}{\sqrt{j_2}})= \\
\!\!\!\!&=\!\!\!\!&\frac{1}{P^4_1P^4_2}\Bigl(\langle\phi_1\otimes\phi_2\vert
  {\bf B}_{12}\vert\phi_1\otimes\phi_2\rangle\Bigr)^2+O(\frac{1}{\sqrt{j_1}})
  +O(\frac{1}{\sqrt{j_2}}),
\end{eqnarray*}
which yields that the second term in the brackets in (\ref{eq:4.3.6}) is 
also vanishing in the $j_1,j_2\to\infty$ limit. 

Therefore, as a summary of the results of the above calculations, we have 
proven the following statement: 

\begin{theorem*}
Let $L_1,\cdots,L_N$ be straight lines in $\mathbb{R}^3$ such that no two of 
them are parallel. Then there are $E(3)$-invariant elementary quantum 
mechanical systems ${\cal S}_1,...,{\cal S}_N$ and a sequence of their pure 
quantum states $\phi_{1k}$,...,$\phi_{Nk}$, $k\in\mathbb{N}$, indexed by 
pairs $(s_{1k},P_{1k})$,...,$(s_{Nk},P_{Nk})$ of their Casimir invariants, 
respectively, such that, in the $(\vert s_1\vert,P_1),...,(\vert s_N\vert,
P_N)\to\infty$ limit, the magnitude $\vert d_{\bi\bj}\vert$ of the empirical 
distances tend with asymptotically vanishing uncertainty to the Euclidean 
distances $D_{\bi\bj}$ between the straight lines $L_{\bi}$ and $L_{\bj}$, given 
by (\ref{eq:2.2.4}), for any ${\bi,\bj}=1,...,N$. 
\end{theorem*}

\noindent
Thus, the metric structure of the Euclidean 3-space could be recovered in 
the classical limit from appropriate quantum observables of Euclidean 
invariant elementary quantum mechanical systems. 

Following the same strategy, the calculation of the classical limit of the 
empirical angles and 3-volume elements is quite straightforward: the 
$m_1=j_1=s_1$, $m_2=j_2=s_2\to\infty$ limit of the empirical angles $\omega
_{12}$, given by (\ref{eq:4.3.4}), is $\beta_{12}$; and the analogous limit 
of the 3-volume element given by (\ref{eq:4.3.5}) is just the Euclidean 
3-volume element. These results, together with the Theorem above, provide 
an extension of the Spin Geometry Theorem of Penrose \cite{Pe79,Pe} from 
$SU(2)$ to $E(3)$-invariant systems. 


\section{Final remarks}
\label{sec-5}

The relative position vectors $d^a_{\bi\bj}$, ${\bi},{\bj}=1,2,3$, of a 
three-particle system also make it possible to define a notion of angle 
that is \emph{different} from that we considered in subsections \ref{sub-2.3} 
and \ref{sub-4.3}: 
\begin{equation*}
\cos\varpi_{12,32}:=\frac{\delta_{ab}d^a_{12}d^b_{32}}{\vert d_{12}\vert\,\vert 
d_{32}\vert}=\frac{\delta_{ab}(\varepsilon^a{}_{cd}p^c_1p^d_2)(\varepsilon^b
{}_{ef}p^e_3p^f_2)}{\vert\varepsilon^a{}_{bc}p^b_1p^c_2\vert\,\vert\varepsilon^a
{}_{bc}p^b_3p^c_2\vert}=\frac{P^2_2P^2_{13}-P^2_{12}P^2_{23}}{\sqrt{P^2_1P^2_2-
P^4_{12}}\sqrt{P^2_2P^2_3-P^4_{23}}} \label{eq:5.1}
\end{equation*}
with range $\varpi_{12,32}\in[0,\pi]$ defines the angle \emph{between the 
relative position vectors} pointing from the 
second subsystem's centre-of-mass line to that of the first and the third 
subsystems, respectively. The angles $\varpi_{23,13}$ and $\varpi_{21,31}$ are 
defined analogously. Although at the classical level this angle coincides 
with the Euclidean one, the analogous empirical angle in the quantum 
theory deviates from $\omega_{\bi\bj}$. Another concept of the empirical 
3-volume element could also be introduced, as the volume of the tetrahedron 
spanned by the three vectors $\varepsilon^a{}_{bc}p^b_{\bi}p^c_{\bj}$. Thus, at 
the fundamental, quantum level there might not exist unique, \emph{a priori} 
obvious analog of the classical geometrical notions, like angle, distance or 
3-volume element. In addition to the requirement of their correct behaviour 
in the classical limit (and their `naturalness' and `usefulness'), can we 
have some selection rule to choose one from the various possibilities? 

The states of the composite system by means of which the correct classical 
limit of the various empirical geometrical quantities could be derived are 
\emph{pure tensor product states}, built from the pure vector states of the 
elementary subsystems. Thus, in deriving these, we did \emph{not} need to 
use entangled states of the composite system. But then, if the subsystems are 
independent, how can one obtain the distance, angle, etc. between them? The 
answer is that the entanglement of the subsystems can be considered to be 
\emph{already built into the structure of the observables of the composite 
system}. In fact, the operators by means of which the empirical distances, 
angles and 3-volume elements are defined have the structure ${\bf W}_{12}=
\frac{1}{2}\delta_{ab}({\bf p}^a_1{\bf J}^b_2+{\bf J}^a_1{\bf p}^b_2)$, 
${\bf P}^2_{12}=\delta_{ab}{\bf p}^a_1{\bf p}^b_2$ and $\varepsilon_{abc}{\bf p}
^a_1{\bf p}^b_2{\bf p}^c_3$, respectively. These observables of the 
\emph{composite} system are `entanglements' of the \emph{observables of the 
subsystems}. The \emph{states} of the subsystems do not need to be entangled. 

As we already noted in subsection \ref{sub-4.2}, if the state of the 
composite system is mixed or entangled, then the interpretation of the 
empirical distance (and of the angles and 3-volume elements, too) is not 
obvious, and e.g. the distance between two subsystems may depend on the 
state of other subsystems. In this case, the resulting distances cannot be 
expected to be compatible with the structure of any metric space. Thus, by 
assuming that the state of the composite system is a pure tensor product 
state we implicitly assumed that the subsystems are independent, and hence 
interact with one another weakly. If, however, the subsystems are inextricably 
entangled (e.g. since they are very strongly interacting with one another), 
then the `quantum geometry' defined by such systems may not be expected even 
to resemble to the Euclidean geometry \emph{at all}. The Euclidean structure 
of the classical `physical 3-space' that we see appears to be defined only 
by the independent, very weakly interacting subsystems of the Universe. 


\section{Acknowledgments}
\label{sec-6}

Thanks are due to J\"org Frauendiener and Paul Tod for their useful remarks 
on the determination of the inverse Clebsch--Gordan coefficients; to 
L\'aszl\'o Feh\'er for the clarification of the algebraic structures on the 
dual of $e(3)$; to P\'eter Vecserny\'es for the numerous discussions on the 
algebraic formulation of quantum theory; and to Ted Jacobson for the 
discussion on the role of entanglement in the emergence of the geometry of 
the 3-space/spacetime from quantum mechanics. 

No funds, grants or support was received.


\appendix

\section{Appendix}
\label{sec-A}
\subsection{Complex coordinates and the line bundles ${\cal O}(-2s)$
over ${\cal S}_P$}
\label{sub-A.1}

In the complex stereographic coordinates $(\zeta,\bar\zeta)$ on ${\cal S}_P$, 
defined by $\zeta:=\exp({\rm i}\varphi)\cot(\theta/2)$ in terms of the 
familiar spherical polar coordinates $(\theta,\varphi)$, the Cartesian 
components of the `position vector' $p^a$ in the classical momentum space 
and the complex null tangent $m^a$, respectively, are 
\begin{equation}
p^a=P\Bigl(\frac{\bar\zeta+\zeta}{1+\zeta\bar\zeta},{\rm i}\frac{\bar\zeta-
  \zeta}{1+\zeta\bar\zeta},\frac{\zeta\bar\zeta-1}{1+\zeta\bar\zeta}\Bigr),
\hskip 20pt
m^a=\frac{1}{\sqrt{2}}\Bigl(\frac{1-\zeta^2}{1+\zeta\bar\zeta},{\rm i}
  \frac{1+\zeta^2}{1+\zeta\bar\zeta},\frac{2\zeta}{1+\zeta\bar\zeta}\Bigr). 
\label{eq:A.1.1}
\end{equation}
These imply that $p^a\varepsilon_{abc}=-{\rm i}P(m_b\bar m_c-\bar m_bm_c)$, 
where $\bar m_a$ is the complex conjugate of $m_a$. Also in these coordinates, 
the line element of the metric and the corresponding area element on 
${\cal S}_P$, respectively, are 
\begin{equation}
dh^2=\frac{4P^2}{(1+\zeta\bar\zeta)^2}d\zeta d\bar\zeta, \hskip 20pt 
{\rm d}{\cal S}_P=-\frac{2{\rm i}P^2}{(1+\zeta\bar\zeta)^2}d\zeta\wedge d
\bar\zeta. \label{eq:A.1.2}
\end{equation}
These are just the metric and area element inherited from the metric and 
volume element of momentum 3-space, respectively. The complex null vectors 
$m^a$ and $\bar m^a$ are unique up to a phase as they are $(1,0)$ and $(0,1)$ 
type vectors, respectively, in the natural complex structure of ${\cal S}_P
\approx S^2$ (see e.g. \cite{HT}). As a differential operator, $m^a$ is given 
by 
\begin{equation}
m^a\bigl(\frac{\partial}{\partial p^a}\bigr)=\frac{1}{\sqrt{2}P}\bigl(1+
\zeta\bar\zeta\bigr)\bigl(\frac{\partial}{\partial\bar\zeta}\bigr). 
\label{eq:A.1.3}
\end{equation}
The contraction of the complex null vectors, $m^a$ and $\bar m^a$, as well 
as of the `position vector' $p^a$ in the momentum space with the Pauli 
matrices can also be expressed by the vectors $\{o^A,\iota^A\}$ of the 
(normalized) Newman--Penrose spinor basis: $m^a\sigma^{AA'}_a=-o^A\bar\iota
^{A'}$, $\bar m^a\sigma^{AA'}_a=-\iota^A\bar o^{A'}$ and $p^a\sigma^{AA'}_a=P
(\iota^A\bar\iota^{A'}-o^A\bar o^{A'})/\sqrt{2}$. The components of the 
vectors of the Newman--Penrose basis in the spinor basis $\{O^A,I^A\}$ 
associated with the Cartesian vector basis are 
\begin{equation*}
o^A=\frac{-{\rm i}}{\sqrt{1+\zeta\bar\zeta}}\left(\begin{array}{c}
     \zeta \\
     1 \\
     \end{array}\right), 
\hskip 20pt
\iota^A=\frac{-{\rm i}}{\sqrt{1+\zeta\bar\zeta}}
     \left(\begin{array}{c}
     1 \\
     -\bar\zeta \\
     \end{array}\right).
\end{equation*}
Note that this basis is well defined only on ${\cal S}_P$ minus its `north 
pole', which is the domain of the coordinate system $(\zeta,\bar\zeta)$, as 
well as of the complex null vectors $m^a$ and $\bar m^a$. 

A scalar $\phi$ is said to have spin weight $s=\frac{1}{2}(p-q)$ if under 
the rescaling $\{o^A,\iota^A\}\mapsto\{\lambda o^A,\lambda^{-1}\iota^A\}$, 
$\lambda\in\mathbb{C}-\{0\}$, the scalar $\phi$ transforms as $\phi\mapsto
\lambda^p\bar\lambda^q\phi$ (see \cite{HT,PR1}). The bundle of such scalars 
is denoted by ${\cal O}(-2s)$. The complex line bundle ${\cal O}(-2s)$ is 
globally trivializable precisely when $s=0$; otherwise it has a twist. 
The domain of the coordinate system $(\zeta,\bar\zeta)$ is a local 
trivialization domain for ${\cal O}(-2s)$ for any $s$. For a detailed 
discussion of the line bundles ${\cal O}(-2s)$, see e.g. \cite{HT,PR1}. 

The edth and edth-prime operators of Newman and Penrose \cite{NP} acting on 
spin weighted functions, e.g. on the cross section $\phi$ of ${\cal O}(-2s)$, 
can be defined by 
\begin{equation}
{\edth}\phi=\frac{1}{\sqrt{2}P}\Bigl(\bigl(1+\zeta\bar\zeta\bigr)
   \frac{\partial\phi}{\partial\bar\zeta}+s\zeta\phi\Bigr), \hskip 20pt
{\edth}'\phi=\frac{1}{\sqrt{2}P}\Bigl(\bigl(1+\zeta\bar\zeta\bigr)
   \frac{\partial\phi}{\partial\zeta}-s\bar\zeta\phi\Bigr); \label{eq:A.1.4}
\end{equation}
and hence for their commutator we obtain that $({\edth}{\edth}'-{\edth}'
{\edth})\phi=-(1/P^2)s\phi$. It is not difficult to check that ${\edth}p^a=
m^a$, ${\edth}m^a=0$ and ${\edth}'m^a=-p^a/P^2$. 

A purely algebraic introduction of the \emph{spin weighted spherical 
harmonics}, given in \cite{PR1}, is based on the comparison of the 
appropriate symmetrized products of the vectors of the Cartesian spinor 
basis $\{O_A,I_A\}$ and those of the Newman--Penrose spinor basis $\{o_A,
\iota_A\}$ adapted to the unit sphere (and given explicitly by equation 
(4.15.98) of \cite{PR1}): 
\begin{equation}
{}_sY_{jm}:=N_{s,j,m}O_{(A_1}\cdots O_{A_{j-m}}I_{A_{j-m+1}}\cdots I_{A_{2j})}
o^{A_1}\cdots o^{A_{j+s}}\iota^{A_{j+s+1}}\cdots\iota^{A_{2j}}, \label{eq:A.1.5}
\end{equation}
where the coefficient $N_{s,j,m}$ is 
\begin{equation}
N_{s,j,m}:=(-)^{j+m}\sqrt{\frac{2j+1}{4\pi}}\frac{(2j)!}{\sqrt{(j-m)!
(j+m)!(j-s)!(j+s)!}}, \label{eq:A.1.6}
\end{equation}
and $2s\in\mathbb{Z}$, $j=\vert s\vert,\vert s\vert+1,\vert s\vert+2,...$ 
and $m=-j,-j+1,...,j$. This choice of the normalization factor yields that 
the spherical harmonics ${}_0Y_{jm}$ coincide with the standard expressions 
for the ordinary spherical harmonics $Y_{jm}$. The action of the edth 
operators on the harmonics ${}_sY_{jm}$ is 
\begin{equation}
{\edth}{}_sY_{jm}=-\frac{1}{\sqrt{2}P}\sqrt{(j+s+1)(j-s)}\,{}_{s+1}Y_{jm}, 
\hskip 10pt
{\edth}'{}_sY_{jm}=\frac{1}{\sqrt{2}P}\sqrt{(j-s+1)(j+s)}\,{}_{s-1}Y_{jm}.
\label{eq:A.1.7}
\end{equation}
The harmonics ${}_sY_{j\,m}$ form an orthonormal basis in the space of the 
spin weighted functions with spin weight $s$ with respect to the $L_2$ 
scalar product on the \emph{unit} 2-sphere (see e.g. \cite{PR1}). 

A spin weighted scalar $\phi$ is called \emph{holomorphic} if ${\edth}'\phi
=0$, and \emph{anti-holomorphic} if ${\edth}\phi=0$. It is known (see e.g. 
\cite{HT,PR1}) that $\dim\ker{\edth}=\dim\ker{\edth}'=0$ for $s<0$ and $s>0$, 
respectively; and $\dim\ker{\edth}=\dim\ker{\edth}'=2\vert s\vert+1$ for $s
\geq0$ and $s\leq0$, respectively. By (\ref{eq:A.1.7}) these kernels are 
spanned by the special spherical harmonics ${}_sY_{\vert s\vert\,m}$. Hence the 
spin weight of the \emph{holomorphic} cross sections is non-positive, $s=
-\vert s\vert$, while that of the \emph{anti-holomorphic} ones is 
non-negative, $s=\vert s\vert$. They form $2\vert s\vert+1$ dimensional 
subspaces in ${\cal H}_{P,s}$. 


\subsection{The evaluation of $\langle{}_sY_{kn}\vert{\bf p}_a\vert\,{}_s
Y_{jm}\rangle$ and $\langle{}_sY_{kn}\vert{\bf J}_a\vert\,{}_sY_{jm}
\rangle$}
\label{sub-A.2}

(\ref{eq:A.1.5}) implies that in the product of two spherical harmonics, 
${}_{s_1}Y_{j_1m_1}\,{}_{s_2}Y_{j_2m_2}$, the difference of the total number of the 
$I_A$ and of the $O_A$ spinors is $2(m_1+m_2)$, and the difference of the total 
number of the $o^A$ and of the $\iota^A$ spinors is $2(s_1+s_2)$. Hence, the 
spin weight of ${}_{s_1}Y_{j_1m_1}\,{}_{s_2}Y_{j_2m_2}$ is $s_1+s_2$, and in its 
expansion in terms of spin weighted spherical harmonics only the harmonics of 
the form ${}_{s_1+s_2}Y_{j(m_1+m_2)}$ appear. Thus, there are constants $C(s_1,
j_1,m_1;s_2,j_2,m_2\vert j)$ such that 
\begin{equation}
{}_{s_1}Y_{j_1m_1}\,{}_{s_2}Y_{j_2m_2}=\sum_j C\bigl(s_1,j_1,m_1;s_2,j_2,m_2
\vert j\bigr)\,{}_{s_1+s_2}Y_{j(m_1+m_2)}, \label{eq:A.2.1}
\end{equation}
where, as one can show, $\max\{\vert j_1-j_2\vert,\vert s_1+s_2\vert\}\leq j
\leq j_1+j_2$. These constants are analogous to the (inverse) of the usual 
Clebsch--Gordan coefficients, and hence these may also be called the 
(inverse) C-G coefficients. 

By (\ref{eq:A.1.1}) and the explicit expression of the ordinary spherical 
harmonics in the coordinates $(\zeta,\bar\zeta)$, the components of the 
linear momentum are 
\begin{equation}
p_a=P\sqrt{\frac{2\pi}{3}}\Bigl({}_0Y_{1-1}-{}_0Y_{11},{\rm i}({}_0Y_{1-1}+
{}_0Y_{11}),\sqrt{2}\,{}_0Y_{10}\Bigr). \label{eq:A.1.1a}
\end{equation}
Hence, to determine the matrix elements $\langle{}_sY_{kn}\vert{\bf p}_a\vert
{}_sY_{jm}\rangle$, we need to calculate the expansion (\ref{eq:A.2.1}) only 
for ${}_0Y_{1n}\,{}_sY_{jm}$. 

This calculation is based on (\ref{eq:A.1.5}), in which, following \cite{PR1}, 
we introduce the notations $Z(j,m)_{A_1...A_{2j}}:= O_{(A_1}\cdots O_{A_{j-m}}
I_{A_{j-m+1}}\cdots I_{A_{2j})}$ and ${}_sZ(j,m):=(N_{s,j,m})^{-1}{}_sY_{jm}$. Then 
for any $M_A$ and any totally symmetric spinor $Z_{A_1...A_{2j}}$ the complete 
algebraically irreducible decomposition of their 
product is 
\begin{equation*}
Z_{A_1...A_{2j}}M_A=Z_{(A_1...A_{2j}}M_{A)}+\frac{1}{2j+1}\varepsilon_{A_1A}
M^BZ_{BA_2...A_{2j}}+\cdots+\frac{1}{2j+1}\varepsilon_{A_{2j}A}M^BZ_{A_1...
A_{2j-1}B}.
\end{equation*}
Applying this formula to $I_A$ and $Z(j,m)_{A_1...A_{2j}}$, we obtain 
\begin{eqnarray}
Z(j,m)_{A_1...A_{2j}}I_A\!\!\!\!&=\!\!\!\!&Z(j+\frac{1}{2},m+\frac{1}{2})
 _{AA_1...A_{2j}}+ \label{eq:A.2.2a} \\
&{}&+\frac{1}{2j+1}\varepsilon_{A_1A}I^BO_{(B}O_{A_2}\cdots O_{A_{j-m}}
  I_{A_{j-m+1}}\cdots I_{A_{2j})}+\cdots + \nonumber\\
&{}&+\frac{1}{2j+1}\varepsilon_{A_{j-m}A}I^BO_{(A_1}\cdots O_{A_{j-m-1}}O_B
  I_{A_{j-m+1}}\cdots I_{A_{2j})}= \nonumber\\
\!\!\!\!&=\!\!\!\!&Z(j+\frac{1}{2},m+\frac{1}{2})_{AA_1...A_{2j}}+ \nonumber\\
&{}&+\frac{1}{(2j+1)2j}\Bigl\{(j-m)\varepsilon_{A_1A}O_{(A_2}\cdots O_{A
  _{j-m}}I_{A_{j-m+1}}\cdots I_{A_{2j})}+\cdots+ \nonumber\\
&{}&+(j-m)\varepsilon_{A_{2j}A}
  O_{(A_1}\cdots O_{A_{j-m-1}}I_{A_{j-m}}\cdots I_{A_{2j-1})}\Bigr\}= \nonumber
  \\
\!\!\!\!&=\!\!\!\!&Z(j+\frac{1}{2},m+\frac{1}{2})_{AA_1...A_{2j}}-
  \frac{j-m}{2j+1}\varepsilon_{A(A_1}Z(j-\frac{1}{2},m+\frac{1}{2})
  _{A_2...A_{2j})}. \nonumber
\end{eqnarray}
In a similar way 
\begin{equation}
Z(j,m)_{A_1...A_{2j}}O_A=Z(j+\frac{1}{2},m-\frac{1}{2})_{AA_1...A_{2j}}+\frac{j+m}
{2j+1}\varepsilon_{A(A_1}Z(j-\frac{1}{2},m-\frac{1}{2})_{A_2...A_{2j})}. 
\label{eq:A.2.2b}
\end{equation}
These two are the key formulae on which the present calculation of the 
(inverse) C-G coefficients is based. In particular, using the technique of 
complete algebraic irreducible decomposition of the various spinors, the 
repeated application of these formulae yields 
\begin{eqnarray}
Z(j,m)_{A_1...A_{2j}}\!\!\!\!&{}\!\!\!\!&I_AI_B=Z(j+1,m+1)_{ABA_1...A_{2j}}- 
  \nonumber\\
-\!\!\!\!&{}\!\!\!\!&\frac{j-m}{2(j+1)}\Bigl(\varepsilon_{A(A_1}Z(j,m+1)
  _{A_2...A_{2j})B}+\varepsilon_{B(A_1}Z(j,m+1)_{A_2...A_{2j})A}\Bigr)- \nonumber\\
-\!\!\!\!&{}\!\!\!\!&\frac{(j-m)(j-m-1)}{2j(2j+1)}\varepsilon_{A(A_1}Z(j-1,m+1)
  _{A_2...A_{2j-1}}\varepsilon_{A_{2j})B}, \label{eq:A.2.3a}\\
Z(j,m)_{A_1...A_{2j}}\!\!\!\!&{}\!\!\!\!&I_AO_B=Z(j+1,m)_{ABA_1...A_{2j}}-\frac{1}{2}
  \varepsilon_{AB}Z(j,m)_{A_1...A_{2j}}+ \nonumber\\
+\!\!\!\!&{}\!\!\!\!&\frac{m}{2(j+1)}\Bigl(\varepsilon_{A(A_1}Z(j,m)
  _{A_2...A_{2j})B}+\varepsilon_{B(A_1}Z(j,m)_{A_2...A_{2j})A}\Bigr)+ \nonumber\\
+\!\!\!\!&{}\!\!\!\!&\frac{(j-m)(j+m)}{2j(2j+1)}\varepsilon_{A(A_1}Z(j-1,m)
  _{A_2...A_{2j-1}}\varepsilon_{A_{2j})B}, \label{eq:A.2.3b}\\
Z(j,m)_{A_1...A_{2j}}\!\!\!\!&{}\!\!\!\!&O_AO_B=Z(j+1,m-1)_{ABA_1...A_{2j}}+ 
  \nonumber\\
+\!\!\!\!&{}\!\!\!\!&\frac{j+m}{2(j+1)}\Bigl(\varepsilon_{A(A_1}Z(j,m-1)
  _{A_2...A_{2j})B}+\varepsilon_{B(A_1}Z(j,m-1)_{A_2...A_{2j})A}\Bigr)- \nonumber\\
-\!\!\!\!&{}\!\!\!\!&\frac{(j+m)(j+m-1)}{2j(2j+1)}\varepsilon_{A(A_1}Z(j-1,m-1)
  _{A_2...A_{2j-1}}\varepsilon_{A_{2j})B}. \label{eq:A.2.3c}
\end{eqnarray}
Since $Z(\frac{1}{2},\frac{1}{2})_A=I_A$, $Z(\frac{1}{2},-\frac{1}{2})_A=O_A$, 
$Z(1,1)_{AB}=I_AI_B$, $Z(1,0)_{AB}=I_{(A}O_{B)}$ and $Z(1,-1)_{AB}$ $=O_AO_B$, 
equations (\ref{eq:A.2.2a})-(\ref{eq:A.2.3c}), together with (\ref{eq:A.1.5}) 
and (\ref{eq:A.1.6}), already give 
\begin{eqnarray}
{}_0Y_{1,1}\,{}_sY_{j,m}\!\!\!\!&=\!\!\!\!&\sqrt{\frac{3}{8\pi}}\frac{1}{j+1}
  \sqrt{\frac{(j-s+1)(j+s+1)(j+m+1)(j+m+2)}{(2j+1)(2j+3)}}{}_sY_{j+1,m+1}-
  \nonumber\\
\!\!\!\!&-\!\!\!\!&\sqrt{\frac{3}{8\pi}}\frac{s}{j(j+1)}\sqrt{(j+m+1)(j-m)}
  {}_sY_{j,m+1}- \nonumber\\
\!\!\!\!&-\!\!\!\!&\sqrt{\frac{3}{8\pi}}\frac{1}{j}\sqrt{\frac{(j-s)(j+s)
  (j-m-1)(j-m)}{(2j-1)(2j+1)}}{}_sY_{j-1,m+1}; \label{eq:A.2.4a} \\
{}_0Y_{1,0}\,{}_sY_{j,m}\!\!\!\!&=\!\!\!\!&\sqrt{\frac{3}{4\pi}}\frac{1}{j+1}
  \sqrt{\frac{(j+1+s)(j+1-s)(j+1+m)(j+1-m)}{(2j+1)(2j+3)}}{}_sY_{j+1,m}+
  \nonumber\\
\!\!\!\!&+\!\!\!\!&\sqrt{\frac{3}{4\pi}}\frac{sm}{j(j+1)}{}_sY_{j,m}+ 
  \nonumber\\
\!\!\!\!&+\!\!\!\!&\sqrt{\frac{3}{4\pi}}\frac{1}{j}\sqrt{\frac{(j-s)(j+s)
  (j+m)(j-m)}{(2j-1)(2j+1)}}{}_sY_{j-1,m}; \label{eq:A.2.4b} \\
{}_0Y_{1,-1}\,{}_sY_{j,m}\!\!\!\!&=\!\!\!\!&\sqrt{\frac{3}{8\pi}}\frac{1}{j+1}
  \sqrt{\frac{(j+1+s)(j+1-s)(j-m+1)(j-m+2)}{(2j+1)(2j+3)}}{}_sY_{j+1,m-1}+
  \nonumber\\
\!\!\!\!&+\!\!\!\!&\sqrt{\frac{3}{8\pi}}\frac{s}{j(j+1)}\sqrt{(j+m)(j-m+1)}
  {}_sY_{j,m-1}- \nonumber\\
\!\!\!\!&-\!\!\!\!&\sqrt{\frac{3}{8\pi}}\frac{1}{j}\sqrt{\frac{(j+s)(j-s)
  (j+m)(j+m-1)}{(2j-1)(2j+1)}}{}_sY_{j-1,m-1}; \label{eq:A.2.4c}
\end{eqnarray}
where recall that, on the right hand sides, the spherical harmonics ${}_s
Y_{j,m}$ are vanishing if $j<\vert s\vert$ or $j<\vert m\vert$. (In these 
formulae, to avoid confusion, we inserted a comma between the indices $j$ 
and $m$ of ${}_sY_{jm}$.) 

Then, using (\ref{eq:A.1.1a}) and (\ref{eq:A.2.4a})-(\ref{eq:A.2.4c}), we 
find that the only non-zero matrix elements of ${\bf p}_a$ are 
\begin{eqnarray}
\langle{}_sY_{j+1,n}\vert{\bf p}_1\vert{}_sY_{j,m}\rangle\!\!\!\!&{}\!\!\!\!&=
  \frac{P^3}{2(j+1)}\sqrt{\frac{(j+s+1)(j-s+1)}{(2j+1)(2j+3)}}
  \label{eq:A.2.5a} \\
\times\!\!\!\!&{}\!\!\!\!&\Bigl(\sqrt{(j-m+1)(j-m+2)}\delta_{n,m-1}-
  \sqrt{(j+m+1)(j+m+2)}\delta_{n,m+1}\Bigr), \nonumber\\
\langle{}_sY_{j,n}\vert{\bf p}_1\vert{}_sY_{j,m}\rangle\!\!\!\!&{}\!\!\!\!&=
  \frac{P^3s}{2j(j+1)} \label{eq:A.2.5b}\\
\times\!\!\!\!&{}\!\!\!\!&\Bigl(\sqrt{(j+m)(j-m+1)}\delta_{n,m-1}+\sqrt{(j-m)
  (j+m+1)}\delta_{n,m+1}\Bigr), \nonumber\\
\langle{}_sY_{j-1,n}\vert{\bf p}_1\vert{}_sY_{j,m}\rangle\!\!\!\!&{}\!\!\!\!&=
  \frac{P^3}{2j}\sqrt{\frac{(j+s)(j-s)}{(2j-1)(2j+1)}}\label{eq:A.2.5c}\\
\times\!\!\!\!&{}\!\!\!\!&\Bigl(\sqrt{(j-m)(j-m-1)}\delta_{n,m+1}-\sqrt{(j+m)
  (j+m-1)}\delta_{n,m-1}\Bigr); \nonumber
\end{eqnarray}
\begin{eqnarray}
\langle{}_sY_{j+1,n}\vert{\bf p}_2\vert{}_sY_{j,m}\rangle\!\!\!\!&{}\!\!\!\!&=
  {\rm i}\frac{P^3}{2(j+1)}\sqrt{\frac{(j+s+1)(j-s+1)}{(2j+1)(2j+3)}}
  \label{eq:A.2.6a} \\
\times\!\!\!\!&{}\!\!\!\!&\Bigl(\sqrt{(j+m+1)(j+m+2)}\delta_{n,m+1}+
  \sqrt{(j-m+1)(j-m+2)}\delta_{n,m-1}\Bigr), \nonumber\\
\langle{}_sY_{j,n}\vert{\bf p}_2\vert{}_sY_{j,m}\rangle\!\!\!\!&{}\!\!\!\!&=
  {\rm i}\frac{P^3s}{2j(j+1)} \label{eq:A.2.6b}\\
\times\!\!\!\!&{}\!\!\!\!&\Bigl(\sqrt{(j+m)(j-m+1)}\delta_{n,m-1}-
  \sqrt{(j-m)(j+m+1)}\delta_{n,m+1}\Bigr), \nonumber\\
\langle{}_sY_{j-1,n}\vert{\bf p}_2\vert{}_sY_{j,m}\rangle\!\!\!\!&{}\!\!\!\!&=
  -{\rm i}\frac{P^3}{2j}\sqrt{\frac{(j+s)(j-s)}{(2j-1)(2j+1)}}
  \label{eq:A.2.6c}\\
\times\!\!\!\!&{}\!\!\!\!&\Bigl(\sqrt{(j+m)(j+m-1)}\delta_{n,m-1}+
  \sqrt{(j-m)(j-m-1)}\delta_{n,m+1}\Bigr), \nonumber
\end{eqnarray}
\begin{eqnarray}
\langle{}_sY_{j+1,n}\vert{\bf p}_3\vert{}_sY_{j,m}\rangle\!\!\!\!&=\!\!\!\!&
  \frac{P^3}{j+1}\sqrt{\frac{(j+s+1)(j-s+1)(j+m+1)(j-m+1)}{(2j+1)(2j+3)}}
  \delta_{n,m}, \label{eq:A.2.7a}\\
\langle{}_sY_{j,n}\vert{\bf p}_3\vert{}_sY_{j,m}\rangle\!\!\!\!&=\!\!\!\!&
  \frac{P^3ms}{j(j+1)}\delta_{n,m}, \label{eq:A.2.7b}\\
\langle{}_sY_{j-1,n}\vert{\bf p}_3\vert{}_sY_{j,m}\rangle\!\!\!\!&=\!\!\!\!&
  \frac{P^3}{j}\sqrt{\frac{(j+s)(j-s)(j+m)(j-m)}{(2j-1)(2j+1)}}\delta_{n,m}. 
  \label{eq:A.2.7c}
\end{eqnarray}
Thus, the subspaces spanned by ${}_sY_{j,m}$ with given $s$ and $j$ are not 
invariant under the action of the momentum operators; and while ${\bf p}_3$ 
does not change the index $m$, ${\bf p}_1\pm{\rm i}{\bf p}_2$ 
increases/decreases the value of $m$. 

Next, we calculate the matrix elements of the angular momentum operator 
using $m^a={\edth}p^a$, $\bar m^a={\edth}'p^a$, equations (\ref{eq:A.1.7}) 
and the expression (\ref{eq:3.1.5b}) for the angular momentum vector 
operator. By integration by parts we obtain 
\begin{eqnarray}
\langle{}_sY_{k,n}\vert{\bf J}_a\vert{}_sY_{j,m}\rangle\!\!\!\!&=\!\!\!\!&
  \hbar P\int_{{\cal S}_P}\overline{{}_sY_{k,n}}\Bigl(({\edth}p_a)({\edth}'{}_s
  Y_{j,m})-({\edth}'p_a)({\edth}{}_sY_{j,m})+s\frac{p_a}{P^2}{}_sY_{j,m}\Bigr)
  {\rm d}{\cal S}_P= \nonumber \\
\!\!\!\!&=\!\!\!\!&\hbar P\int_{{\cal S}_P}\Bigl(-(\overline{{\edth}'{}_s
  Y_{k,n}})p_a({\edth}'{}_sY_{j,m})+(\overline{{\edth}{}_sY_{k,n}})p_a({\edth}
  {}_sY_{j,m})+ \nonumber \\
\!\!\!\!&{}\!\!\!\!&\hskip 35pt +\overline{{}_sY_{k,n}}p_a\bigl({\edth}'
  {\edth}{}_sY_{j,m}-{\edth}{\edth}'{}_sY_{j,m}\bigr)+s\frac{p_a}{P^2}
  \overline{{}_sY_{k,n}}\,{}_sY_{j,m}\Bigr){\rm d}{\cal S}_P =  \nonumber\\
\!\!\!\!&=\!\!\!\!&\frac{\hbar}{2P}\Bigl\{\sqrt{(k+s+1)(k-s)}\sqrt{(j+s+1)
  (j-s)}\langle{}_{s+1}Y_{k,n}\vert{\bf p}_a\vert{}_{s+1}Y_{j,m}\rangle-
  \nonumber\\
\!\!\!\!&{}\!\!\!\!&\hskip 20pt -\sqrt{(k-s+1)(k+s)}\sqrt{(j-s+1)(j+s)}
  \langle{}_{s-1}Y_{k,n}\vert{\bf p}_a\vert{}_{s-1}Y_{j,m}\rangle+\nonumber\\
\!\!\!\!&{}\!\!\!\!&\hskip 20pt +4s\langle{}_sY_{k,n}\vert{\bf p}_a\vert{}_s
  Y_{j,m}\rangle\Bigr\}. \label{eq:A.2.8}
\end{eqnarray}
Hence, the matrix elements of the angular momentum vector operator are 
simple expressions of those of the linear momentum. Using 
(\ref{eq:A.2.5a})-(\ref{eq:A.2.7c}) and (\ref{eq:A.2.8}), we find that the 
only non-zero matrix elements of ${\bf J}_a$ are 
\begin{eqnarray}
\langle{}_sY_{j,n}\vert{\bf J}_1\vert{}_sY_{j,m}\rangle=\frac{1}{2}\hbar
  \!\!\!\!&{P^2}\!\!\!\!&\Bigl(\sqrt{(j+m)(j-m+1)}\delta_{n,m-1}+ \nonumber \\
\!\!\!\!&{}\!\!\!\!&+\sqrt{(j-m)(j+m+1)}\delta_{n,m+1}\Bigr), 
  \label{eq:A.2.9a} \\
\langle{}_sY_{j,n}\vert{\bf J}_2\vert{}_sY_{j,m}\rangle=\frac{\rm i}{2}
\hbar\!\!\!\!&{P^2}\!\!\!\!&\Bigl(\sqrt{(j+m)(j-m+1)}\delta_{n,m-1}-
  \nonumber \\
\!\!\!\!&{}\!\!\!\!&-\sqrt{(j-m)(j+m+1)}\delta_{n,m+1}\Bigr), 
  \label{eq:A.2.9b} \\
\langle{}_sY_{j,n}\vert{\bf J}_3\vert{}_sY_{j,m}\rangle=\hbar\!\!\!\!&{P^2}
  \!\!\!\!& m\delta_{n,m}. \label{eq:A.2.9c} 
\end{eqnarray}
These are precisely the well known matrix elements of the angular momentum 
operator in quantum mechanics. In particular, these are independent of the 
spin weight of the spherical harmonics. 


\end{document}